\documentclass[12pt]{iopart}

\usepackage{color}
\usepackage{graphicx}
\usepackage{amssymb}

\def\be{\begin{equation}}
\def\ee{\end{equation}}
\def\ba{\begin{array}}
\def\ea{\end{array}}
\def\beq{\begin{eqnarray}}
\def\eeq{\end{eqnarray}}

\begin{document}

\title{Charging and energy fluctuations of a driven quantum battery}

\author{A Crescente$^{1,2}$, M Carrega$^{3}$, M Sassetti$^{1,2}$ and D Ferraro$^{1,2}$}

\address{$^{1}$Dipartimento di Fisica, Universit\`a di Genova, Via Dodecaneso 33, 16146, Genova, Italy}
\address{$^{2}$SPIN-CNR, Via Dodecaneso 33, 16146 Genova, Italy}
\address{$^{3}$NEST, Istituto Nanoscienze-CNR and Scuola Normale Superiore, Piazza S. Silvestro 12, I-56127 Pisa, Italy}


\begin{abstract}
We consider a quantum battery modeled as a set of $N$ independent two-level quantum systems driven by a time dependent classical source.
Different figures of merit, such as stored energy, time of charging and energy quantum fluctuations during the charging process, are characterized in a wide range of parameters, by means of numerical approach and suitable analytical approximation scheme.
Particular emphasis is put on the role of different initial conditions, describing the preparation state of the quantum battery, as well as on the sensitivity to the functional form of the external time-dependent drive. 
It is shown that an optimal charging protocol, characterized by fast charging time and the absence of charging fluctuations, can be achieved  starting from the ground state of each two-level system, while other pure preparation states are less efficient.
Moreover, we argue that a periodic train of peaked rectangular pulses can lead to fast charging. This study aims at providing a useful theoretical background in view of future experimental solid-state implementations. 
\end{abstract}

\section{Introduction}

One of the main task in modern technology is to find smart ways to  exploit quantum resources to realize new devices able to outperform their classical counterpart~\cite{Riedel17, Acin18, QT_China, QT_USA, QT_Canada}. In this framework, increasing interest is devoted to the development of quantum technologies for energy storage and power supply. Here, thermodynamic concepts have been investigated in a quantum setting~\cite{Esposito09,Vinjanampathy16,Bera19, Depasquale18, Carrega19, Benenti17, Pekola15, Levy12} by means of several methods, initially inspired also by quantum information theory~\cite{DiVincenzo95}.

It has been argued that quantum systems can be used to store energy, thus coining the word ``quantum batteries'' (QBs)~\cite{Campaioli18}, and that it is possible to enhance charging power (energy stored in a given time interval) and work extraction performances by taking advantage of quantum correlations~\cite{Alicki13, Hovhannisyan13, Binder15, Campaioli17}.
Several directions are currently under study, both to demonstrate the quantum advantage of single or many-body QBs~\cite{Binder15, Le18, Julia-Farre18, Rossini19, Rosa19}, and both to propose realistic models of QBs in view of solid-state implementations~\cite{Ferraro18}.
They include e.g. arrays of superconducting qubits~\cite{Devoret13} or quantum dots~\cite{Singha11} realized in semiconducting nanostructures. Each individual building block of these systems can be effectively described in terms of a quantum Two-Level System (TLS)~\cite{Wiel02, Koch07}.
The charging of these single cell of QBs can be achieved for example by properly controlling in time the direct spin-spin like interaction between different cells~\cite{Campaioli18, Le18, Rossini19, Rosa19} or the effective interaction between TLSs induced by the coupling with an external cavity radiation~\cite{Ferraro18, Andolina18, Andolina19, Ferraro19}. In these models, the interplay between the collective behavior associated to the presence of various interacting TLSs and their quantum features can lead to a faster energy storage (higher average charging power) with respect to their non-interacting counterparts~\cite{Julia-Farre18, Rossini19, Rosa19, Andolina19b}.
Despite these very interesting predictions, such models and the necessary non-local or collective behaviours~\cite{Julia-Farre18, Rossini19} are difficult to be implemented in actual experiments~\cite{Hofheinz08, Fink09}.

A simpler possible way to implement a QB consists in a collection of qubits (TLSs) whose charging dynamics is controlled by means of a constant (static) external bias. 
However, this can lead to an efficient and fast charging only for very strong values of the external signal. This limitation can be overcome by considering a time dependent external classical drive. A first analysis of such possibility has been recently carried out in Refs.~\cite{Zhang19, Chen19}. There, authors considered the performance associated to the charging of a TLS due to a harmonic drive focussing on a particular subset of parameters.

Another crucial, and still largely unexplored, aspect of the dynamics of QBs is related to energy quantum fluctuations~\cite{Rosa19, Hamma19}. Indeed, strong fluctuations of stored energy would undermine the effective working of a QB, leading to incomplete charging and reduction of the average charging power~\cite{Hamma19}, namely the ratio between the stored energy and the minimal time needed to achieve complete charging~\cite{Binder15, Campaioli17}.

In the present paper, we analyze a driven QB subjected to an ac field, along three main directions. First of all, we investigate the energy transfer between the classical source and the QB in order to identify parameter regimes where the charging can occur faster and more efficiently with respect to a static case. Moreover, a complete characterization is carried out by discussing the performance of the charging with respect to different initial states of the TLS. We also consider the role of the functional form of the external drive, by focussing on two specific examples, namely a monochromatic source and a train of rectangular pulses, showing that in the latter case speed up in the charging process can be achieved.
To characterize energy quantum fluctuations during the charging of a QB, two independent energy correlators are considered, namely fluctuations at equal time or between the initial and final instant of the charging process \cite{Friis17}, for different preparation states and driving shapes.
This analysis provides useful information for future actual  implementations of solid-state based QBs, with detailed description of non-interacting QBs subjected to ac external drives, which can be further extended including interactions among single cells.

The paper is organized as follows. In Section~\ref{Model} we introduce the model for a QB coupled to a classical external time-dependent source describing both the numerical approach and the analytical approximation used to solve it. In Section~\ref{Initial} the different initial conditions of the TLS and the shape of the external drives are presented, while in Section~\ref{Relevant} we introduce the main figures of merit we will analyze. We report the results of our study in Section~\ref{Results}. Finally Section~\ref{Conclusions} is devoted to the conclusions. We report examples of other possible initial states (including mixed states) in \ref{app_a}, while some useful expressions for the time evolution of the pure initial states are reported in \ref{app_b} and asymptotic limits for the charging energy are reported in \ref{app_c}. 


\section{Model and general settings}
\label{Model}

\begin{figure}[htbp]
\centering
\includegraphics[width=0.95 \textwidth]{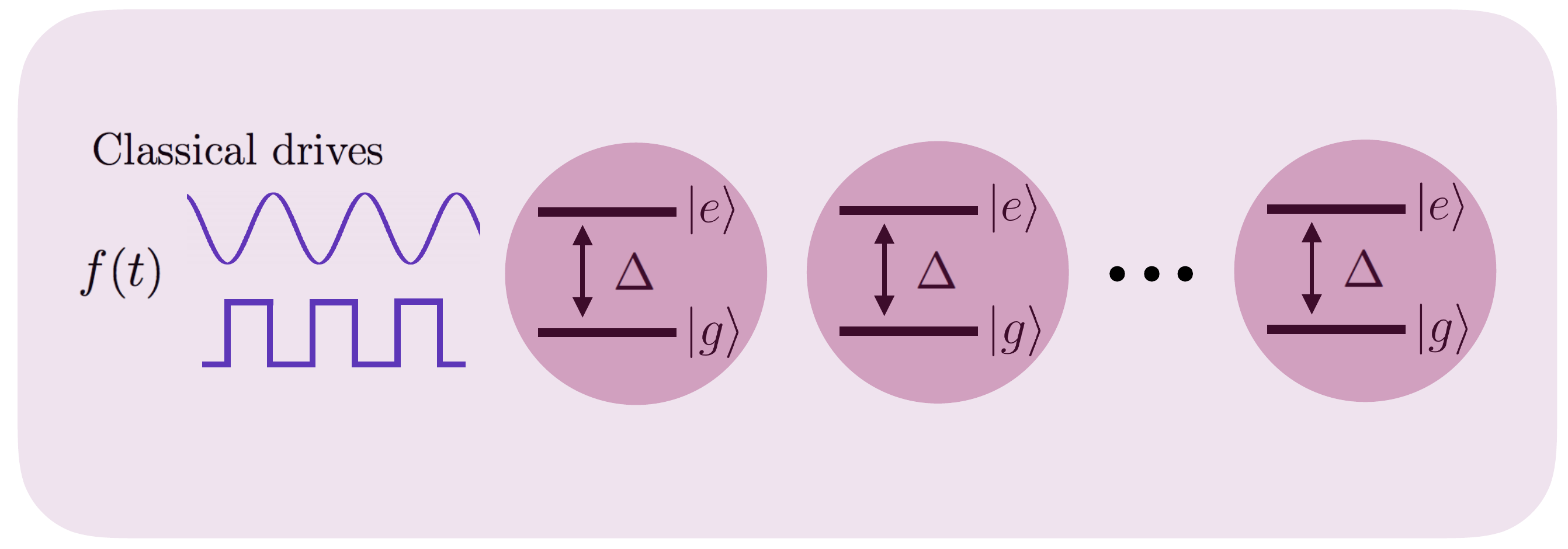}
\caption{Scheme of a QB composed of $N$ independent TLSs. Charging is achieved by means of an external time dependent drive $f(t)$. The single cells of the battery are assumed identical, with an energy gap $\Delta$ between the ground and the excited state.}
\label{Figure1}
\end{figure}

We consider a QB modelled as a set of $N$ independent cells schematized as TLSs\footnote{Notice that this assumption allows us to properly explore the single cell dynamics under periodic driving, avoiding additional contributions related to many-body effects and non-classical correlations among the cells that may influence the performance of the device and crucially depend on the form of the interaction~\cite{Campaioli18, Le18, Rossini19, Rosa19,Ferraro18, Andolina18, Andolina19, Ferraro19,Lu12}.}

All single cells are assumed to be identical and coupled to the same external time-dependent classical force, as sketched in Figure~\ref{Figure1}. The external force then acts as a charger for the whole QB.
The dynamics of the system during the charging process is described by the Hamiltonian (hereafter we set $\hbar =1$)
\be
\label{htot}
\hat{\mathcal{H}}_{{\rm tot}} (t)=\sum_{i=1}^N \hat{\mathcal{H}}^{(i)}(t)= \sum_{i=1}^N \left(\hat{\mathcal{H}}_0^{(i)} + \hat{\mathcal{H}}^{(i)}_1 (t)\right),
\ee
where $\hat{\mathcal{H}}_0^{(i)}$ describes the $i$-th time independent TLS  of the QB, while $\hat{\mathcal{H}}_1^{(i)}(t)$ represents the coupling with the external drive.

Since all single cells are independent and interact separately with the charger, the total energy has a linear scaling with $N$, being an estensive quantity. Therefore, in the following, we can focus on the charging dynamics and performances of a single TLS, omitting the index $i$ for notational convenience.
We thus have
\begin{equation}
\label{H} 
\hat{\mathcal{H}}(t)=\hat{\mathcal{H}}_{0} +\hat{\mathcal{H}}_{1}(t)= \frac{\Delta}{2}\hat{\sigma}_{z}+\frac{A}{2} \Theta(t) f(t) \hat{\sigma}_{x},
\end{equation}
where $\Delta$ is the level spacing between the ground state $|g\rangle$ and the excited state $|e\rangle$ of the TLS and $\hat{\sigma}_{i}$ $(i=x,y,z)$ are the usual Pauli matrices. In the above Equation, the last term describes the coupling with a classical drive, assuming that it is switched on at time $t=0$ (see the step function $\Theta(t)$), with  $A$ the amplitude of the external time dependent drive $f(t)$. In the following we will focus on purely ac periodic drives with period $T=2\pi/\omega$ and null average 
\begin{equation}
\label{Condition1} 
\frac{1}{T}\int_0^T dt' f(t')=0 
\end{equation}
and satisfying the normalization condition
\begin{equation}
\frac{1}{T}\int_0^T  dt' |f(t')|^{2}=1.
\end{equation}
The TLS is initially (for $t<0$) prepared in a given state (see below) and, for $t\geq 0$, it evolves with the whole Hamiltonian $\hat{\mathcal{H}}(t)$ for a given time interval $t_c$, after which the drive is switched off. 

We stress that here we focus on the dynamics of a closed quantum system. Possible interactions with the external environment leads to relaxation and dephasing characterized by typical time scales $t_{r}$ and $t_{\varphi}$ respectively which depend on the actual physical implementation of the TLS~\cite{Haroche_Book, Uli_book}. In general, dissipation can strongly affect the dynamics of a quantum system \cite{Uli_book, Makhlin01} and therefore one may expect that also charging performances of a QB will be influenced, especially in the case of strong coupling with the reservoir. However, for practical implementations the possibility to accurately control and mitigate dissipation effects has been demonstrated, achieving the limit of very weak coupling with the environment and consequent very long values of $t_r$ and $t_\varphi$ \cite{Devoret13, Wendin17}. Therefore, in the following we will restrict our analysis to evolution times $t $ such that $t\ll t_{r}, t_{\varphi}$ where dissipation effects can be safely neglected.

\subsection{Numerical solution}
The full dynamics of the system can be solved numerically, starting from the Schr\"odinger equation
\begin{equation}
\label{Schroedinger}
i\frac{d|\psi(t)\rangle}{dt}=\hat{\mathcal{H}}(t)|\psi(t)\rangle
\end{equation}
with $|\psi(t)\rangle$ describing the two-component spinor wave-function of the TLS at a given time $t$.

Considering Equation (\ref{H}), one obtains the set of coupled differential equations 
\begin{eqnarray}
  \frac{d \sigma_{x}(t)}{dt} &=& \Delta \sigma_{y}(t),   \label{exact-eq11} \\
  \frac{d \sigma_{y}(t)}{dt} &=&  -\Delta \sigma_{x}(t) + A f(t)  \sigma_{z}(t),   \label{exact-eq22} \\
  \frac{d \sigma_{z}(t)}{dt} &=&  -A f(t) \sigma_{y}(t),  \label{exact-eq33}
\end{eqnarray}
where we have introduced the quantities
\begin{eqnarray}
\sigma_{x}(t)&=&\langle \psi(t)|\hat{\sigma}_x | \psi(t)\rangle,\\ 
\sigma_{y}(t)&=&\langle \psi(t)|\hat{\sigma}_y|\psi(t)\rangle,\\ 
\sigma_{z}(t)&=&\langle \psi(t)|\hat{\sigma}_z|\psi(t)\rangle
\end{eqnarray}
components of the vector that describes the state of the TLS on the Bloch sphere \cite{Haroche_Book}.
The numerical solution of this system of coupled differential equations provides the full dynamic of the QB in a wide parameter range, as we will discuss below.

\subsection{Useful approximation scheme}
\label{CHRW_section}
An useful approach to shed light on the physical mechanisms behind the dynamics of this system is the so-called Counter-Rotating Hybridized Rotating-Wave (CHRW) approximation~\cite{Lu12, Lu16, Yan17}. This extends the conventional rotating wave scheme by taking into account, at least partially, counter-rotating terms which can play relevant role in presence of time-dependent external drives.
Here, we briefly recall this method whose implementation is based on the following steps.

First, one considers the time dependent rotation 
\begin{equation}\label{U(t)}
\hat{\mathcal{U}}(t)= \exp{\left[i\pi z \varphi(t) \hat{\sigma}_{x}\right]}
\end{equation}
with $z\in \mathbb{R}$ a free dimensionless parameter whose value will be fixed shortly, and 
\begin{equation}\label{varphi}
\varphi(t)= \int_0^t \frac{dt'}{T} f(t').
\end{equation}
The Hamiltonian responsible for the time evolution is then 
\begin{eqnarray}
\hat{\mathcal{H}}'&=& \hat{\mathcal{U}}\hat{\mathcal{H}}\hat{\mathcal{U}}^\dagger-i\hat{\mathcal{U}}\frac{d}{dt}\hat{\mathcal{U}}^\dagger \\
&=& \frac{\Delta}{2} \bigg\{ \cos\left[2\pi z \varphi(t)\right]\hat{\sigma}_{z} + \sin\left[2\pi z \varphi(t)\right]\hat{\sigma}_{y}\bigg\}+ \frac{1}{2}\left(A-z\omega \right)  f(t)\hat{\sigma}_{x},
\end{eqnarray}
where we have used the identity

\begin{equation} e^{i\gamma \sigma_j}= \cos(\gamma)\mathbb{I} + i\sin(\gamma)\sigma_j, \end{equation}
with $j=x,y,z$ labelling the Pauli matrices $\sigma_j$.
Using the Fourier series decomposition
\begin{equation}
\begin{array}{rcl}
\cos\left[2 \pi z \varphi(t)\right]&=&p_0(z) + \sum_{l=1}^{\infty}\cos(l\omega t) [p_l(z)+ p_{-l}(z)], \\
\sin\left[2 \pi z \varphi(t)\right]&=&\sum_{l=1}^{\infty}\sin(l\omega t) [p_l(z)- p_{-l}(z)],\\
f(t)&=&2\sum_{k=1}^{\infty} c_k \cos (\omega k t)
\end{array}
\end{equation}
$\hat{\mathcal{H}}'$ can then be written as 
\begin{eqnarray}
\label{H_series} 
\hat{\mathcal{H}}'&=& \frac{\Delta}{2} \bigg\{ \bigg[p_0(z) +\sum_{l=1}^\infty \cos(l\omega t)\bigg(p_l(z)+p_{-l}(z)\bigg)\bigg]\hat{\sigma}_z \nonumber \\
&+& \bigg[\sum_{l=1}^\infty \sin(l\omega t)\bigg(p_l(z)-p_{-l}(z)\bigg)\bigg]\hat{\sigma} _y\bigg\}\nonumber\\
&+&\left(A-z\omega \right)  \sum_{k=1}^{\infty} c_k \cos (\omega k t)\hat{ \sigma}_x. \nonumber
\end{eqnarray}
Notice that, since the considered drives have null average, one has $c_0=0$.

In the previous expressions we have indicated with
\begin{equation}
\label{p_l}
p_l(z) = \int_{-\frac{T}{2}}^{+\frac{T}{2}} \frac{dt}{T}e^{i\omega l t}e^{-i2\pi z \varphi(t)}
\end{equation}
the photoassisted coefficients of the external drive. 
 In order to implement the CHRW approximation one can divide $\hat{\mathcal{H}}'$ into three distinct contributions, namely: 
\begin{eqnarray}
\label{CHRW}
\hat{\mathcal{H}}'_{0}&=& \frac{\Delta}{2} p_0(z)\hat{\sigma}_z \\
\hat{\mathcal{H}}'_{1}&=&\frac{\Delta}{2} \bigg\{\bigg[\bigg(p_1(z)-p_{-1}(z)\bigg)\sin(\omega t)\bigg]\hat{\sigma}_y\bigg\} + \left(A-z\omega \right) c \cos(\omega t)\hat{\sigma}_x\\
\label{first_line}
&=& \bigg(\frac{e^{i \omega t}S_++e^{-i \omega t}S_-}{2}\bigg)\bigg[\bigg(A-z\omega\bigg)c-\frac{\Delta}{2} \bigg(p_1(z)-p_{-1}(z)\bigg)\bigg]+\\
&+& \bigg(\frac{e^{i \omega t}S_-+e^{-i \omega t}S_+}{2}\bigg)\bigg[\bigg(A-z\omega\bigg)c+\frac{\Delta}{2} \bigg(p_1(z)-p_{-1}(z)\bigg)\bigg]\nonumber\\
\hat{\mathcal{H}}'_{2}&=& \hat{\mathcal{H}}'-\hat{\mathcal{H}}'_{0}-\hat{\mathcal{H}}'_{1}
\end{eqnarray}
where we have denoted $c_{1}\equiv c $ for notational convenience and we have introduced 
\begin{equation}
\hat{S}_{\pm}= \frac{1}{2}\left(\hat{\sigma}_{x}\pm i \hat{\sigma}_{y}\right).
\end{equation}

In essence the CHRW approximation consists in neglecting the Hamiltonian contribution $\hat{\mathcal{H}}'_{2}$, which contains higher-order harmonics of the drive,
and forcing the value $z=\bar{z}$ in such a way to cancel the counter-rotating terms, namely the first line in Equation (\ref{first_line}), satisfying the transcendental equation 
\begin{equation}
\label{bar_chi}
 \left(A-\bar{z}\omega\right) c-\frac{\Delta}{2} \bigg[p_1\left(\bar{z}\right)-p_{-1}\left(\bar{z}\right)\bigg]=0.  
 \end{equation}
We thus arrive at
\begin{eqnarray} 
\label{H_approx}
\hat{\mathcal{H}}''&=& \hat{\mathcal{H}}_{0}'(\bar{z})+\hat{\mathcal{H}}'_{1}(\bar{z})\\
&=&  \frac{\Delta}{2}p_0\left(\bar{z}\right)\hat{\sigma}_z + \tilde{A}(e^{i\omega t}\hat{S}_-+e^{-i \omega t}\hat{S}_+), 
\end{eqnarray}
where 
\begin{equation}\label{atilde}
\tilde{A}=  \left(A-\bar{z}\omega\right) c.
\end{equation} 

Finally, by using the unitary transformation
\begin{equation}\label{S(t)}
\hat{\mathcal{S}}(t)=\exp\left[i\frac{\omega t}{2}\hat{\sigma} _z\right]
\end{equation}
we can recast the above Hamiltonian in a time-independent form as
\begin{eqnarray} 
\label{H_final}
\tilde{H}&=& \hat{\mathcal{S}}\hat{\mathcal{H}}''\hat{\mathcal{S}}^\dagger -i \hat{\mathcal{S}} \frac{d}{dt}\hat{\mathcal{S}}^\dagger\nonumber \\
&=& \frac{\tilde{\Delta}}{2} \hat{\sigma}_z+\tilde{A}\hat{\sigma}_x
\end{eqnarray}
with 
\begin{equation}\label{dtilde}
\tilde{\Delta}= \Delta p_0\left(\bar{z}\right)- \omega.
\end{equation}
It is worth to note that Equation (\ref{H_final}) closely resembles the effective Hamiltonian obtained in the conventional rotating wave approximation~\cite{Haroche_Book}, but with renormalized parameters that take into account also the effects of counter-rotating terms, and it has thus a wider range of validity (see below). 

\section{Initial conditions and driving shapes}
\label{Initial}
It is well known that the dynamics of a closed system strongly depends on the initial conditions. Moreover, also the knowledge of the precise shape of the external drives is needed to properly describe the evolution of the QB. We now discuss these two aspects.
\subsection{Pure initial states}\label{pis}
We recall that the initial state of each TLS can be described in terms of the density matrix
\begin{equation}
\label{rho_zero}
\rho_0=\rho(t=0)=\left(\begin{array}{cc}
1-P & \alpha-i\beta\\
\alpha+i\beta & P
\end{array}\right)
\end{equation}
with $0\leq P\leq 1$, $\alpha$ and $\beta$ real coefficients satisfying $\alpha^2+\beta^2\leq P\left( 1-P\right)$ since the trace should be unitary $Tr\left[\rho_{0}\right]=1$ and positive definite $Tr\left[\rho_{0}^2\right]\leq 1$, the equality holding for a pure state. We will focus on three different initial conditions for the QB, corresponding to three different pure states of the TLS. Our choice of initial pure states is dictated by the fact that in a generic mixture with $\alpha=\beta=0$ and $P\neq0,1$ (including thermal states) the full charging of the battery cannot be achieved  as will be clearer in the following~\cite{Alicki13, Campaioli17, Andolina18, Friis17}. 

First, we consider the system to be in the ground state $|g\rangle$ of the TLS ($P=1$, $\alpha=\beta=0$), corresponding to an empty single cell.
The other two states we analyze are
\begin{equation}
|p_x\rangle = \frac{1}{\sqrt{2}} \left(|g\rangle + |e\rangle\right),
\label{eq:px}
\end{equation}
with $P=\alpha=1/2$ and $\beta=0$ and
\begin{equation}
|p_y\rangle = \frac{1}{\sqrt{2}}\left(|g\rangle - i |e\rangle\right)
\end{equation}
with $P=\beta=1/2$ and $\alpha=0$. Notice that the dynamics of all these pure states, in absence of relaxation processes, stays confined to the surface of the Bloch sphere~\cite{Haroche_Book, Uli_book}. We will consider other pure and mixed states in \ref{app_a} to strengthen the above considerations.

To find the dynamics of $\sigma_{z}(t)$ in the CHRW approximation we have to consider the time evolution of the different initial states reported in Equations (\ref{timeev})-(\ref{phi_py}) of \ref{app_b}. By doing so we can write $\sigma_{z}(t)$ starting from the ground state
\begin{eqnarray}
\label{Z_g}
\sigma_{z,
g}(t)&=&\langle \psi_{g}(t) |\hat{\sigma}_{z}|\psi_{g}(t)\rangle \nonumber \\
&=&\frac{1-\cos(\Omega_R t)}{\Omega_R^2}\bigg\{4\tilde{A}^2\cos\left[2 \pi \bar{z} \varphi(t)\right]-2\tilde{A}\tilde{\Delta}\sin\left[2\pi \bar{z} \varphi(t)\right]\sin(\omega t)\bigg\}\nonumber \\ 
&-&\cos\left[2\pi \bar{z} \varphi(t)\right]+\frac{2\tilde{A}}{\Omega_R}\sin(\Omega_R t)\sin\left[2 \pi \bar{z} \varphi(t)\right]\cos(\omega t).
\end{eqnarray}

Starting from $|p_{x}\rangle$ one has 
\begin{eqnarray}
\label{Z_px}
\sigma_{z,p_{x}}(t)&=& \langle \psi_{p_{x}}(t) |\hat{\sigma}_{z}|\psi_{p_{x}}(t)\rangle\nonumber \\
&=&\frac{1-\cos(\Omega_R t)}{\Omega_R^2}\bigg\{-\tilde{\Delta}^2\sin\left[2 \pi \bar{z} \varphi(t)\right]\sin\left(\omega t \right) + 2\tilde{A}\tilde{\Delta}\cos\left[2 \pi \bar{z} \varphi (t)\right]\bigg\} \nonumber \\
&+& \frac{\tilde{\Delta}}{\Omega_R}\sin\left(\Omega_{R} t\right)\sin\left[2\pi \bar{z} \varphi(t)\right]\cos\left(\omega t\right) + \sin\left[2\pi \bar{z} \varphi(t)\right]\sin\left(\omega t\right).
\end{eqnarray}

Finally, choosing $|p_{y}\rangle$ as initial condition, one obtains   
\begin{eqnarray}
\label{Z_py}
\sigma_{z,p_{y}}(t)&=& \langle \psi_{p_{y}}(t) |\hat{\sigma}_{z}|\psi_{p_{y}}(t)\rangle \nonumber \\
&=&\frac{\sin\left( \Omega_R t\right)}{\Omega_R}\left\{2\tilde{A}\cos\left[2\pi \bar{z} \varphi(t)\right] -\tilde{\Delta}\sin\left[2\pi \bar{z} \varphi(t)\right] \sin\left(\omega t\right)\right\}\nonumber \\
&+&\cos\left(\Omega_R t\right) \sin\left[2\pi \bar{z} \varphi(t)\right]\cos\left(\omega t\right).
\end{eqnarray}
In the above expressions we have introduced the renormalized Rabi frequency  
\begin{equation}
\label{Rabi}
\Omega_R= \sqrt{\tilde{\Delta}^2+4\tilde{A}^2}.
\end{equation} 
By looking at the above Equations, it is evident that different initial states strongly influence the evolution of $\sigma_{z}(t)$. Moreover, one has that for a generic state described by the density matrix in Equation (\ref{rho_zero}) the condition
\begin{equation}
\sigma_{z}(t)= (2P-1)\sigma_{z, g}(t)+2\alpha \sigma_{z, p_{x}}(t)+2\beta \sigma_{z, p_{y}}(t)
\end{equation}
holds. 
\subsection{Time-dependent drives}
We consider a classical time-dependent source, focussing on two paradigmatic examples of its functional form: a harmonic drive with cosine shape and a train of rectangular pulses. Both shapes have been investigated in several set-ups~\cite{Bauerle18, Iwahori16, Grabert15, Ludovico16, Ferraro14} and can be implemented quite easily in experiments~\cite{Haroche_Book, Bauerle18}. 
The former is given by 
\begin{equation}\label{fcos}
f^{(c)}(t)= \sqrt{2} \cos \omega t
\end{equation}
with 
\begin{equation}
c^{(c)}=\frac{\sqrt{2}}{2}
\end{equation}
and
\begin{equation}
\label{varphi_c}
\varphi^{(c)}(t)= \frac{\sqrt{2}}{2\pi}\sin(\omega t).
\end{equation}
The photoassisted coefficients read 
\begin{equation}
p^{(c)}_l(z)= J_l(\sqrt{2}z)
\end{equation}
with $J_{l}(x)$ Bessel function of $l$-th order~\cite{Gradshteyn_book}.

In the case of a train of rectangular pulses the external drive in the period $[-T/2, T/2]$ can be written as
\begin{eqnarray}\label{frett}
f^{(r)}(t)&=&\sqrt{\frac{1-\eta}{\eta}}\Theta(t)\Theta\left(-t+\eta \frac{T}{2}\right)\nonumber \\
&-&\sqrt{\frac{\eta}{1-\eta}}\Theta\left(t-\eta \frac{T}{2}\right)\Theta\left(-t+T-\eta \frac{T}{2}\right)\nonumber\\
&+&\sqrt{\frac{1-\eta}{\eta}}\Theta\left(t-T+\eta \frac{T}{2}\right)\Theta(-t+T)
\end{eqnarray}
with $\eta$ the width of the pulse. This leads to 
\begin{equation}
c^{(r)}=\frac{1}{\pi}\frac{1}{\sqrt{\eta \left(1-\eta\right)}}
\end{equation}
and
\begin{eqnarray}
\label{varphi_r}
\varphi^{(r)}(t)&=&\frac{t}{T}\sqrt{\frac{1-\eta}{\eta}}\Theta(t)\Theta\left(-t+\eta \frac{T}{2}\right)\nonumber\\
&+&\bigg(\frac{1}{2}-\frac{t}{T}\bigg)\sqrt{\frac{\eta}{1-\eta}}\Theta\left(t-\eta \frac{T}{2}\right)\Theta\left(-t+T-\eta \frac{T}{2}\right)\nonumber \\
&+&\bigg(\frac{t}{T}-1\bigg)\sqrt{\frac{1-\eta}{\eta}}\Theta\left(t-T+\eta \frac{T}{2}\right)\Theta(-t+T).
\end{eqnarray}
In this case the coefficient of the Fourier expansion in (\ref{p_l}) are given by~\cite{Dubois13, Vannucci17, Ferraro18b}
\begin{equation} 
p^{(r)}_l(z)=\frac{\sqrt{\frac{\eta}{1-\eta}}z\sin\left\{\pi[-\sqrt{\eta(1-\eta)}z+\eta l]\right\}}{\pi\left(\sqrt{\frac{\eta}{1-\eta}}z+l\right)\left\{\pi[-\sqrt{\eta(1-\eta)}z+\eta l]\right\}} .
\end{equation}

\section{Figures of merit}
\label{Relevant}
\subsection{Stored energy}
The more natural quantity that characterizes a QB is the energy stored after a given time $t$~\cite{Binder15, Campaioli17, Julia-Farre18,Ferraro18, Andolina18, Andolina19}, namely 
\begin{eqnarray}
\label{Energy} 
E_{\nu}(t)&=& \langle\psi_{\nu}(t)| \hat{\mathcal{H}}_{0}|\psi_{\nu}(t)\rangle - \langle \psi_{\nu}(0)|\hat{\mathcal{H}}_{0}|\psi_{\nu}(0)\rangle\nonumber \\
&=&\frac{\Delta}{2} \left[\sigma_{z, \nu}(t)-\sigma_{z, \nu}(0)\right]
\end{eqnarray}
with $\nu=g, p_{x}, p_{y}$. Typical protocols consider that the external drive is active for a given time interval $0\leq t\leq t_c$ and after it is switched off. This interval needs to be chosen by looking for the optimal value for $t_c$ such that the charging is complete. This can be done by considering the values of times $t_{m}$ at which the maxima of the stored energy occur, namely
\begin{equation}
E(t_{m})=\bar{E}
\end{equation} 
where $\bar{E}=\Delta$ for a TLS starting in the $|g\rangle$ state and $\bar{E}=\Delta/2$ if one chooses $|p_{x}\rangle$ or $|p_{y}\rangle$ as initial states, and finding the minimum among them 
\begin{equation}
t_{c}\equiv \mathrm{min}\left( t_{m}\right).
\end{equation}
 Within the CHRW framework, it is possible to get closed expressions for the {average energy stored} $E(t)$ with different initial conditions (see Equations (\ref{Energy}) and (\ref{Z_g}-\ref{Z_py})).
Starting from the ground state $|g\rangle$, using Equation (\ref{Z_g}) we indeed obtain
\begin{eqnarray}
\label{E_g}
\frac{E_{g}(t)}{\Delta} &=& \frac{1-\cos(\Omega_R t)}{\Omega_R^2}\bigg\{2\tilde{A}^2\cos\left[2\pi \bar{z} \varphi(t)\right]-\tilde{A}\tilde{\Delta}\sin\left[2\pi \bar{z} \varphi(t)\right]\sin(\omega t)\bigg\}\nonumber \\
&-&\frac{1}{2}\cos\left[2\pi \bar{z} \varphi(t)\right]+\frac{\tilde{A}}{\Omega_R}\sin(\Omega_R t)\sin\left[2\pi \bar{z} \varphi(t)\right]\cos(\omega t)+ \frac{1}{2}.
\end{eqnarray}
Instead, considering $|p_x\rangle$ and Equation (\ref{Z_px}) one has
\begin{eqnarray}
\label{E_px}
\frac{E_{p_{x}}(t)}{\Delta} &=& \frac{1-\cos(\Omega_R t)}{\Omega_R^2}\bigg\{-\frac{\tilde{\Delta}^2}{2}\sin\left[2\pi \bar{z} \varphi(t)\right]\sin\left(\omega t\right) + \tilde{A}\tilde{\Delta}\cos\left[2\pi \bar{z} \varphi(t)\right]\bigg\}\nonumber \\
&+& \frac{\tilde{\Delta}}{2\Omega_R}\sin\left(\Omega_{R}t\right)\sin\left[2\pi \bar{z} \varphi(t) \right]\cos\left(\omega t\right) + \frac{1}{2}\sin\left[2\pi \bar{z} \varphi(t)\right]\sin\left(\omega t\right)\nonumber\\
\end{eqnarray}
and for the system prepared in the state $|p_y\rangle$ one obtains
\begin{eqnarray}
\label{E_py}
\frac{E_{p_{y}}(t)}{\Delta} &=& \frac{\sin \left(\Omega_R t\right)}{\Omega_R}\left\{\tilde{A}\cos\left[2\pi \bar{z} \varphi(t)\right] -\frac{1}{2}\tilde{\Delta}\sin\left[2\pi \bar{z} \varphi(t)\right] \sin\left(\omega t\right)\right\}\nonumber\\
&+&\frac{1}{2}\cos \left(\Omega_R t\right)\sin\left[2\pi \bar{z} \varphi(t)\right]\cos\left(\omega t\right).
\end{eqnarray}
In all cases the average energy stored depends on the two different frequencies $\omega$ and $\Omega_{R}$. The competition between them leads to beats in the energy behavior as a function of time. 
From the applicative point of view one is often interested in finding the minimum time needed in order to fully charge the battery maximizing the average charging power ~\cite{Binder15, Campaioli17, Julia-Farre18, Ferraro18, Andolina18, Andolina19}. Values for the charging time $t_c$ for different initial conditions and driving can be extracted numerically and also evaluated in the framework of the CHRW approximation, as reported in \ref{app_c}. We will discuss them in the following.


\subsection{Energy quantum fluctuations}
\label{Fluctuations}
The knowledge of the average energy stored as a function of time as well as the optimal charging time, is not sufficient to fully characterize a QB. Indeed, together with this, it is important to have information about quantum fluctuations of the energy as a function of time. According to initial assumptions, the TLSs composing the QB are identical and independent, therefore neither disorder effects nor many-body fluctuations are present \cite{Rosa19, Hamma19}.
In this respect, one can define the two following independent quantities \cite{Friis17}: the  equal time energy fluctuations 
\begin{equation}
\label{NW}
\Xi_{\rm{tot}}^2(t)= \langle \left[\hat{\mathcal{H}}_{0, \rm{tot}}(t)-\hat{\mathcal{H}}_{0, \rm{tot}}(0)\right]^2\rangle-\left[\langle \hat{\mathcal{H}}_{0, \rm{tot}}(t)-\hat{\mathcal{H}}_{0, \rm{tot}}(0)\rangle\right]^2
\end{equation} 
and the fluctuations between the initial and final time of the charging process
\begin{eqnarray}
\label{NSigma} 
\Sigma^2_{\rm{tot}}(t) = \bigg[\sqrt{\langle \hat{\mathcal{H}}_{0,\rm{tot}}^2(t)\rangle-(\langle \hat{\mathcal{H}}_{0, \rm{tot}}(t)\rangle)^2}-\sqrt{\langle \hat{\mathcal{H}}_{0, \rm{tot}}^2(0)\rangle-(\langle \hat{\mathcal{H}}_{0, \rm{tot}}(0)\rangle)^2}\bigg]^2,\nonumber\\
\end{eqnarray}
with (see Equation (\ref{htot}))
\begin{equation}
\hat{\mathcal{H}}_{0, {\rm tot}} (t)= \sum_{i=1}^N \hat{\mathcal{H}}_0^{(i)}.
\end{equation}
Here, the average is considered with respect to the initial density matrix $\rho_0$ of Equation (\ref{rho_zero}) and we have evaluated the Heisenberg time evolution of the $\hat{\mathcal{H}}_0$ operator.
In complete analogy with the average energy stored also in this case, due to the independence of the QBs, one has 
\begin{eqnarray}
\Xi_{\rm{tot}}^2(t)&=&N\Xi^2(t)\\
\Sigma^2_{\rm{tot}}(t) &=&N\Sigma^2(t)
\end{eqnarray}
with
\begin{equation}
\label{W}
\Xi^2(t)= \langle \left[\hat{\mathcal{H}}_{0}(t)-\hat{\mathcal{H}}_{0}(0)\right]^2\rangle-\left[\langle \hat{\mathcal{H}}_{0}(t)-\hat{\mathcal{H}}_{0}(0)\rangle\right]^2
\end{equation} 
and
\begin{equation}
\label{Sigma} 
\Sigma^2(t) = \bigg[\sqrt{\langle \hat{\mathcal{H}}_{0}^2(t)\rangle-(\langle \hat{\mathcal{H}}_{0}(t)\rangle)^2}-\sqrt{\langle \hat{\mathcal{H}}_0^2(0)\rangle-(\langle \hat{\mathcal{H}}_0(0)\rangle)^2}\bigg]^2
\end{equation}
fluctuations associated to an individual TLS. 

Introducing the short-hand notation for the correlators
\begin{equation}
\label{V} V(t)=\langle \hat{\mathcal{H}}_0^2(t)\rangle-(\langle \hat{\mathcal{H}}_0(t)\rangle)^2 
\end{equation}
and 
\begin{equation}\label{C}
C(t)=\langle \hat{\mathcal{H}}_0(t)\hat{\mathcal{H}}_0(0)\rangle+\langle \hat{\mathcal{H}}_0(0)\hat{\mathcal{H}}_0(t)\rangle -2\langle \hat{\mathcal{H}}_0(t)\rangle\langle \hat{\mathcal{H}}_0(0)\rangle. 
\end{equation}
it is possible to rewrite the above correlators as
\begin{eqnarray}
\label{W_compact}
\Xi^2(t)&=& V(t)+V(0)-C(t)\\
\label{Sigma_compact}
\Sigma^2(t) &=& V(t)+V(0)-2\sqrt{V(t)V(0)},
\end{eqnarray}
which clearly highlight their formal difference. Both correlators can be expressed in terms of the entries of the density matrix (see Equation (\ref{rho_zero})) and, since $\hat{\mathcal{H}}_0(0)$ is proportional to $\hat{\sigma}_z$, of the averaged values of $\hat{\sigma}_{z}$ at a given time $t$ starting from $|g\rangle$, $|p_{x}\rangle$ and $|p_{y}\rangle$ 
 \cite{Grifoni99}, namely
\begin{eqnarray}
\label{sigmagenerale}
V(t)&=& \frac{\Delta^2}{4}\left[1- \sigma_{z}^2(t)\right]\\
\label{C}
C(t)&=&-\frac{\Delta^2}{2} \left[\sigma_{z, g}(t)+(1-2P)\sigma_{z}(t)\right].
\label{cfunction}
\end{eqnarray}
Moreover, we recall that $\sigma_{z, g}(t), \sigma_{z, p_{x}}(t), \sigma_{z,p_{y}}(t)$ satisfy the initial conditions
\begin{eqnarray}
\sigma_{z, g}(0)&=&-1\\ 
\sigma_{z, p_{x}}(0)&=&\sigma_{z, p_{y}}(0)=0.
\label{initialcond}
\end{eqnarray}
To conclude this part, we underline the fact that combining Equations (\ref{Energy}) and (\ref{sigmagenerale}), and considering an initial mixed state with $P\neq0, 1$ and $\alpha=\beta=0$ the average energy stored reduces to
\begin{equation}
E(t)=\frac{\Delta}{2}(2P-1) \left[1+\sigma_{z}(t)\right].
\end{equation} 
This quantity is $E(t)\geq 0$ for $P>1/2$ and $E(t)\leq 0$ for $P<1/2$. According to this, in the former case we can provide energy to the system up to $\Delta(2P-1)<\Delta$, while in the letter we can only extract energy. This further motivates our choice of initial pure states. 

\section{Results and discussions}
\label{Results}
We now present our main numerical results with the goal of finding conditions able to fulfill an (almost) complete and noiseless charging of the QB. A comparison with the analytical approximation will be also provided. 
 Firstly we consider a fixed (equal) choice of drive amplitude and frequency for both cosine and train of rectangular pulses in order to make a fair comparison of the QB performances with different driving shapes and initial conditions.
Subsequently, we identify and discuss the optimal regions in which the different drives work better and we will discuss how the charging time is improved.

\subsection{Average energy stored}
\begin{figure}[htbp]
\centering
\includegraphics[width=0.9 \textwidth]{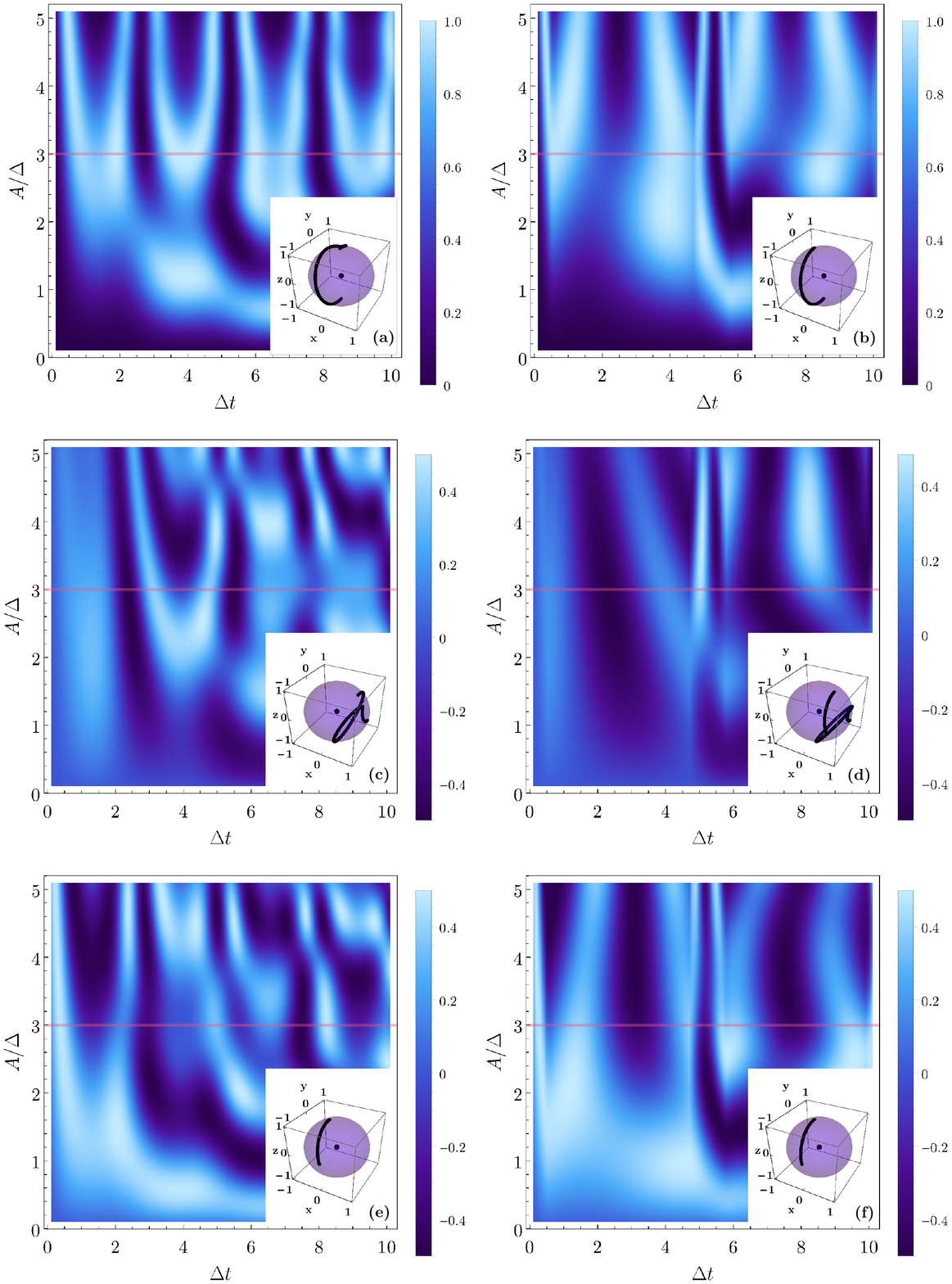}
\caption{Density plot of the average energy stored $E(t)$ (in units of $\Delta$) as a function of $\Delta t$ and $A/\Delta$ and at fixed frequency $\omega=1.2\Delta$. Left panels represent the case of a cosine drive, while the right ones a train of rectangular pulses at $\eta=0.2$ choosing respectively as initial condition $|g\rangle$ a)-b), $|p_{x}\rangle$ c)-d) and $|p_{y}\rangle$ e)-f). Purple horizontal lines indicate the value of $A=3 \Delta$ which is discussed in the text. Insets illustrate the paths followed by the various quantum states on the surface of the Bloch sphere up to the time $t_c$ where the almost complete charging occurs. Parameters associated to these paths are $A=3\Delta$ and $\omega=1.2\Delta$.}
\label{Figure2}
\end{figure}

In Figure \ref{Figure2} we show density plots of the average energy stored $E(t)$ as a function of time and drive amplitude $A$, chosen in the range $0\leq A\leq 5 \Delta$. Bigger values of the drive amplitude are not considered, due to the fact that we want to keep as much confined as possible the power supplied by the external source (related to the drive amplitude). 
Here, we consider a fixed drive frequency $\omega=1.2\Delta$ (close to resonance with the TLS level spacing), different frequencies will be discussed later. We can thus make comparisons between different drives as a function of $A/\Delta$, getting values of the maximum stored energy fairly close to the complete charging of the battery in a time $t_c$ which is rather short compared to $1/\Delta$.
The behaviours of the different initial states of the single cell  in presence of two different shapes of the external drive, cosine and rectangular in Equations (\ref{fcos}) and (\ref{frett}), are reported. 

The overall behaviour of $E(t)$, shown in the density plots, clearly presents beats between the two characteristic frequencies of the system, namely the external drive frequency $\omega$ and the renormalized Rabi frequency $\Omega_{R}$ of Equation (\ref{Rabi}), as already pointed out for the analytic expressions reported in Equations (\ref{E_g}), (\ref{E_px}) and (\ref{E_py}). Moreover, the cosine drive shows a more regular profile with respect to the rectangular pulse as a consequence of the different expression of $\varphi^{(c)}(t)$ with respect to  $\varphi^{(r)}(t)$ (see Equations (\ref{varphi_c}) and (\ref{varphi_r})). 
In all cases, we observe wide regions in the parameter space (white areas) where the charging of the single cell of the QB is almost complete, namely $E\approx \Delta$ starting from the $|g\rangle$ state and  $E\approx \Delta/2$ considering $|p_{x}\rangle$ or $|p_{y}\rangle$ as initial states. 
In presence of time-dependent driving, the full charging is quite easily reached choosing as initial state both the ground state $|g\rangle$ or $|p_{y}\rangle$, while the charging process, starting from $|p_{x}\rangle$, is highly inefficient. Indeed, in the latter case, the regions in the ($A/\Delta$, $\Delta t$) space for which we get the complete charging of the battery is reduced and it takes a longer time $t_c$ to reach it.
This fact can be understood by looking at the paths followed by the state vectors on the surface of the Bloch sphere (see insets of Figure \ref{Figure2}), where we have considered the value of drive amplitude $A=3 \Delta$ represented by the purple cuts on the density plots. 
Indeed, the time dependent drive in Equation (\ref{H}) induces a rotation that occurs mainly around the $\hat{y}$ axis and is able to connect very rapidly both $|g\rangle$ and $|p_{y}\rangle$ to the north pole of the Bloch sphere ($|e\rangle$) corresponding to a complete charging of the single cell of the QB. Very different is the situation concerning the initial state $|p_{x}\rangle$, since it follows a longer and complicated path on the surface of the Bloch sphere.

\begin{figure}[htbp]
\centering
\includegraphics[width=1.02 \textwidth]{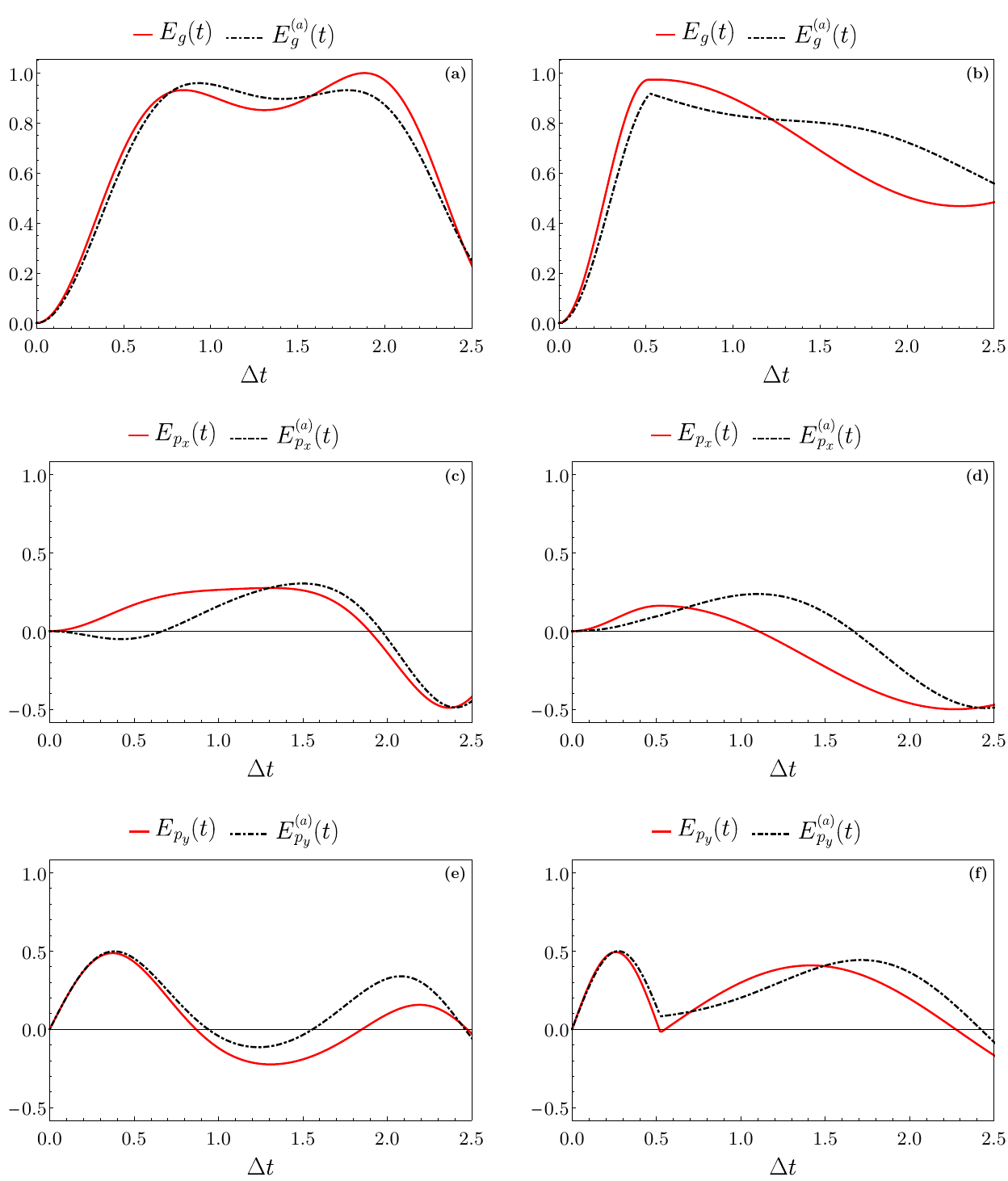}
\caption{Behaviour of $E(t)$ in units of $\Delta$ and as a function of $\Delta t$ for the cosine (left panels) and rectangular drive at $\eta=0.2$ (right panels). Full red curves indicate the numerical results, while dashed-dotted black curves are obtained in the framework of the CHRW approximation. Initial conditions are: $|g\rangle$ a)-b), $|p_{x}\rangle$ c)-d) and $|p_{y}\rangle$ e)-f). Other parameters are $A=3\Delta$ and $\omega=1.2\Delta$.}
\label{Figure3}
\end{figure}

For a more quantitative analysis, we now focus on a representative choice of parameters. We consider $A=3\Delta$ and $\omega=1.2\Delta$ (purple line in Figure \ref{Figure2}), for which the plots of $E(t)$ (in unit of $\Delta$) as a function of $\Delta t$ are reported in Figure \ref{Figure3}. 
Here, examining the full red curves (numerical results), we can find the charging time $ t_c$ obtained where the energy has the maximum $E(t_c)$ closer to the ideal full charge. As a reference we assume the thresholds of almost complete charging processes to be $E \gtrsim 0.9\Delta$ for the ground state and $E \gtrsim 0.45 \Delta$ for the initial state $|p_x\rangle$ and $|p_y\rangle$. 
 The values of the energy and charging time are reported in Table \ref{tab1}. For the ground state of the cosine drive we decided to report data for the first two maxima of the energy because they are comparable in average energy stored, but the first one occurs at roughly half of the time with respect to the second (faster charging).
The charging process for the rectangular drive in the $|g\rangle$ state is faster compared to the cosine drive, but in this regime of parameters reaches a value of the energy which is slightly lower than the second maximum of the cosine. From Table \ref{tab1} we can see that the $|p_y\rangle$ initial condition for both the cosine and rectangular drive lead to an almost complete charging in shorter times, compared to the ground state. Moreover, the charging for a train of rectangular pulses is faster with respect to the one for the cosine shape. Instead, for what it concerns the $|p_x\rangle$ state one obtains the slowest charging times (out of time range shown in Figure \ref{Figure3}) and also in this situation the battery doesn't reach the full charge.
In general we observe that the chosen rectangular pulse appears more efficient. However, by further reducing the width $\eta$ of the rectangular peak (not shown) the system has not enough time to completely charge during the first ramp of the drive, resulting in a detrimental  impact on the performances of the QB.

\begin{table}[!h]
\centering
 \begin{tabular}{ | c | c | c | }
    \hline
     &   Cosine  & Rectangular $(\eta=0.2)$  \\ 
   \cline{2-3} &\begin{tabular}{cc}$E( t_c)$&$\hspace{0.5cm} t_c$\end{tabular}& \begin{tabular}{cc} $E (t_c)$ & $\hspace{0.5cm} t_c$ \end{tabular} \\ \hline
     $|g\rangle$ &\begin{tabular}{cc}$\hspace{0.15cm}0.931$ & $\quad 0.84$ \\ \hline
     $\hspace{0.15cm}0.999$ & $\quad1.88$ \end{tabular} & \begin{tabular}{cc}$\hspace{0.15cm}0.972$ & $\quad0.52$ \end{tabular} \\ \hline 
     $|p_{x}\rangle$&\begin{tabular}{cc} $\hspace{0.15cm}0.389$ & $\quad3.10$\end{tabular} & \begin{tabular}{cc} $\hspace{0.15cm}0.441$ & $\quad5.02$  \end{tabular}\\ \hline 
     $|p_{y}\rangle$ &\begin{tabular}{c c}$\hspace{0.15cm}0.486$ & $\quad0.37$ \end{tabular}& \begin{tabular}{c c}$\hspace{0.15cm}0.493$ & $\quad0.26$ \end{tabular} \\ \hline      
    \end{tabular}
    \caption{Maximum of the stored energy $E(t_c)$ (in units of $\Delta$) and corresponding charging time $t_{c}$ (in units of $1/\Delta$) for the initial conditions $|g\rangle$, $|p_{x}\rangle$ and $|p_{y}\rangle$ for the cosine and rectangular drive ($\eta=0.2$), considered for the red full curves (numerical results) in Figure \ref{Figure3}. Notice that for the ground state of the cosine we have reported both the first and the second maxima.}
    \label{tab1}
\end{table}

We now comment on the regime in which the CHRW approximation, described in Section \ref{CHRW_section}, works well in describing the exact numerical results.
We stress  that both the form of the drive and the initial state play a relevant role in determining the range of validity of the CHRW approximation. In particular, this approximation in the case of a cosine drive holds well both at high enough frequencies ($\omega \gtrsim \Delta$) for arbitrary values of the ratio $A/ \omega$ and at small frequencies ($\omega \lesssim \Delta$) for $A/\omega \lesssim 2$~\cite{Lu12, Lu16, Yan17}. Conversely, one can show that in the case of the train of rectangular pulses at small frequencies ($\omega \lesssim \Delta$) the approximation holds only for $A/\omega\lesssim 1/2$. This is related to the fact that in this case higher order harmonics, neglected in the spirit of the CHRW approximation, play a major role.
Taking into account these conditions, in Figure \ref{Figure3} we show comparison between the exact numerical results obtained so far and the analytical solution achieved within the CHRW approximation in Equations (\ref{Z_g}-\ref{Z_py}) (hereafter denoted with an index $a$) on a short time window $0 \leq \Delta t \leq 2.5$ (within the range of validity discussed above). For the considered range of parameters the condition $\tilde{A} \ll |\tilde{\Delta}|$ (see \ref{app_c}) is well fulfilled. Panel a) shows qualitative agreement between the two approaches for the ground state of the cosine drive. However the value of the maxima and the times at which they occur present a maximal deviation of $\approx 10 \%$.

The $|p_y\rangle$ state shows the best agreement between the numerical and analytical results at short time $\Delta t \lesssim 1$,  and in the other time region shown in Figure \ref{Figure3} e) the curves still present qualitative agreement. In particular the value of the maximum of the energy and the corresponding charging time are almost identical.

Instead, starting from the $|p_x\rangle$ state, the CHRW approach do not reproduce well the behaviour of the average energy stored.
Similar considerations hold true for what it concerns the rectangular drive. Also in this case the $|p_x\rangle$ preparation state behaviour is not reproduced by the CHRW curve, while the $|p_y\rangle$ matches well the value of the maximum and charging time. The qualitative trend of the curves is preserved in all the time region shown.

Furthermore one can show that for the train of rectangular pulses, increasing the value of $\eta$, the agreement between the numerical results and the ones obtained in the CHRW approximation gets progressively better.

\begin{figure}[htbp]
\centering
\includegraphics[width=0.93 \textwidth]{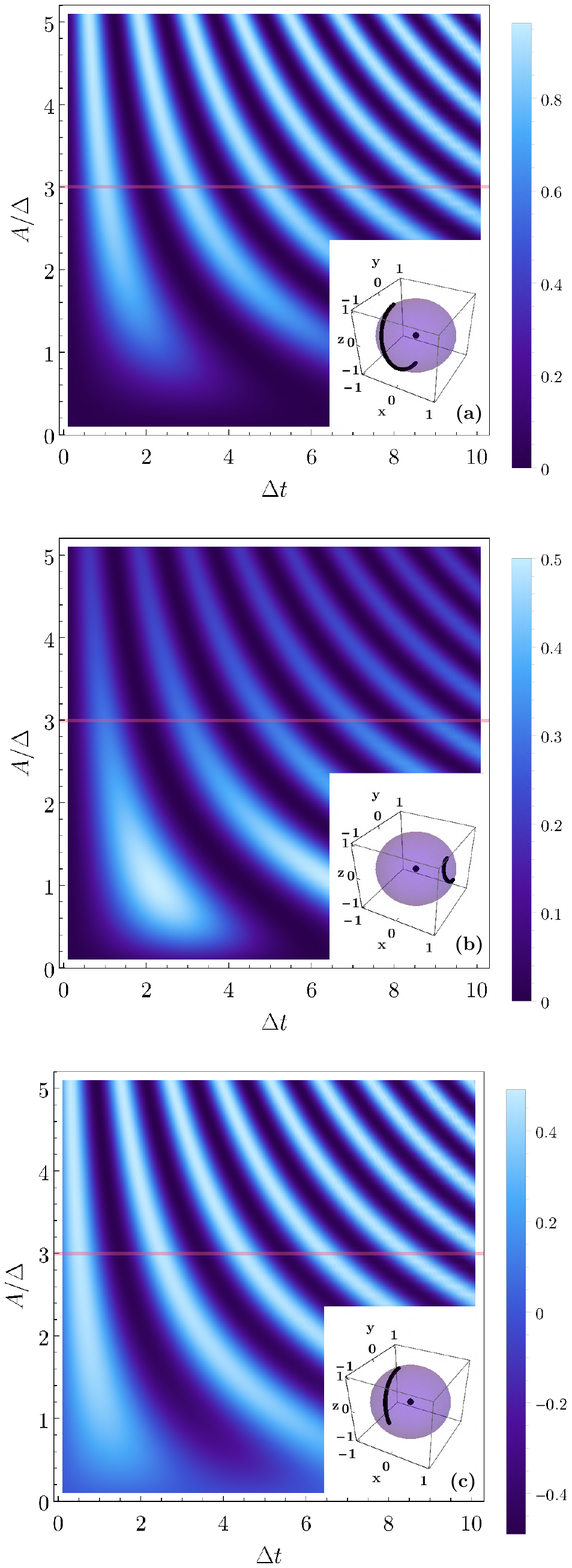}
\caption{Density plot of the average energy stored $E(t)$ (in units of $\Delta$) as a function of $\Delta t$ and $A/\Delta$ for the static case choosing respectively as initial condition $|g\rangle$ a), $|p_{x}\rangle$ b) and $|p_{y}\rangle$ c). Purple horizontal lines indicate the value of $A=3 \Delta$ for which in the insets we illustrate the paths followed by the various quantum states on the surface of the Bloch sphere up to the time $t_c$ where the first maximum of the average energy stored occurs.}\label{Figure4}
\end{figure}

It is worth pointing out that, in the framework of the CHRW approximation, the opposite regime $|\tilde{\Delta}|\ll \tilde{A}$ considered in Ref. \cite{Zhang19} for the harmonic drive case and reported also in \ref{app_c} corresponds to the regions $A\lesssim 0.6 \Delta$ in our density plots. The plots and the above discussions show that, by carefully choosing the initial state and the driving shape, faster charging times can be achieved in the $|\tilde{\Delta}| \gg \tilde{A}$ regime valid for example for the parameters $A=3\Delta$ and $\omega=1.2 \Delta$ we have considered.

To prove the better performance of an ac drive with respect to a static one, in Figure \ref{Figure4} we show the correspondent density plot for the static case (see analytic expressions summarized in Equations (\ref{E_g_static}), (\ref{E_px_static}) and (\ref{E_py_static})). Here, one notices a more regular pattern for $E(t)$ due to the dependence on only one characteristic frequency (the static Rabi frequency $\Omega^{(s)}_{R}=\sqrt{\Delta^{2}+A^{2}}$).
In this case, the evolution of the system along the Bloch sphere follows closed trajectories (circles) in contrast to the ac driven cases where, as long as $\omega$ and $\Omega_{R}$ are incommensurate, the evolution leads to open curves. The maxima approach the values $E\approx \Delta$ for an initial state $|g\rangle$ and $E\approx \Delta/2$ for $|p_{y}\rangle$ only asymptotically (see Equations (\ref{E_g_static}) and (\ref{E_py_static}) in \ref{app_c}) at large values of $A$ with a huge power needed by the source supplier. 
For sake of clarity, we consider the same representative amplitude as before $A=3\Delta$ and we report the results in Table \ref{tab3}. 
 
\begin{table}[!h]
\centering
 \begin{tabular}{ | c | c |}
    \hline
     &  Static    \\ 
  \cline{2-2}  &\begin{tabular}{cc}$E(t_c)$&$\hspace{0.5cm} t_c$\end{tabular} \\ \hline
     $|g\rangle$ &\begin{tabular}{cc}$\hspace{0.15cm}0.830$ & $\quad 1.17$ \end{tabular}  \\ \hline 
     $|p_{x}\rangle$&\begin{tabular}{cc} $\hspace{0.15cm}0.277$ & $\quad 1.01$\end{tabular} \\ \hline 
     $|p_{y}\rangle$ &\begin{tabular}{c c}$\hspace{0.15cm}0.457$ & $\quad0.58$\end{tabular} \\ \hline      
    \end{tabular}
    \caption{Maximum of the stored energy $E(t_c)$ (in units of $\Delta$) and corresponding charging time $ t_{c}$ (in units of $1/\Delta$) for the initial conditions $|g\rangle$, $|p_{x}\rangle$ and $|p_{y}\rangle$ for the static case with $A=3\Delta$ represented by the purple line on the density plot in Figure \ref{Figure4}.}
    \label{tab3}
\end{table} 

Considering the first maximum of the average energy stored, in the static case the charging time starting from  the ground state is longer with respect to the driven ones. This is also true for the $|p_y\rangle$ preparation state.
Starting from the $|p_x\rangle$ state with this choice of parameters, as in the driven case, one cannot fully charge the QB, reaching only a value of $E \approx 0.277 \Delta$ at $\Delta t^{(s)}_{c,p_{x}}\approx1.01$.
However, the density plot in Figure \ref{Figure4}, shows regions where full charging can be achieved at smaller times.
Nevertheless, these charging times are still longer with respect to the ones obtained starting from both the ground and $|p_{y}\rangle$ in the driven cases.
Therefore, for the considered values of drive amplitude and frequency (for the driven case), but also in a wider region of the parameters, as shown in the density plots in Figure \ref{Figure2} and \ref{Figure4}, a static drive is less efficient with respect to the driven cases for the initial states $|g\rangle$ and $|p_{y}\rangle$. 

In the density plots in Figure \ref{Figure2} ($\omega \sim \Delta$) we can see that the $|p_y\rangle$ state wins over $|g\rangle$ and $|p_x\rangle$ at $A\lesssim 2.5\Delta$, while the ground state shows higher average energy compared to the other states for $A\gtrsim 2.5 \Delta$. However, we want to stress that the actual hierarchy among the various initial states crucially depends on the chosen frequency. For $\omega \gtrsim \Delta$ we have that an analogous behaviour occurs at progressively higher value of $A$. Different is the situation for $\omega \lesssim \Delta$, where the $|p_y\rangle$ state is better than the ground state up to $A \sim 5\Delta$. Moreover, for higher values of the drive amplitude they become comparable. 
This clearly emerges from the limiting case of the static drive (see Figure \ref{Figure3}).

\begin{figure}[htbp]
\centering
\includegraphics[width=0.85 \textwidth]{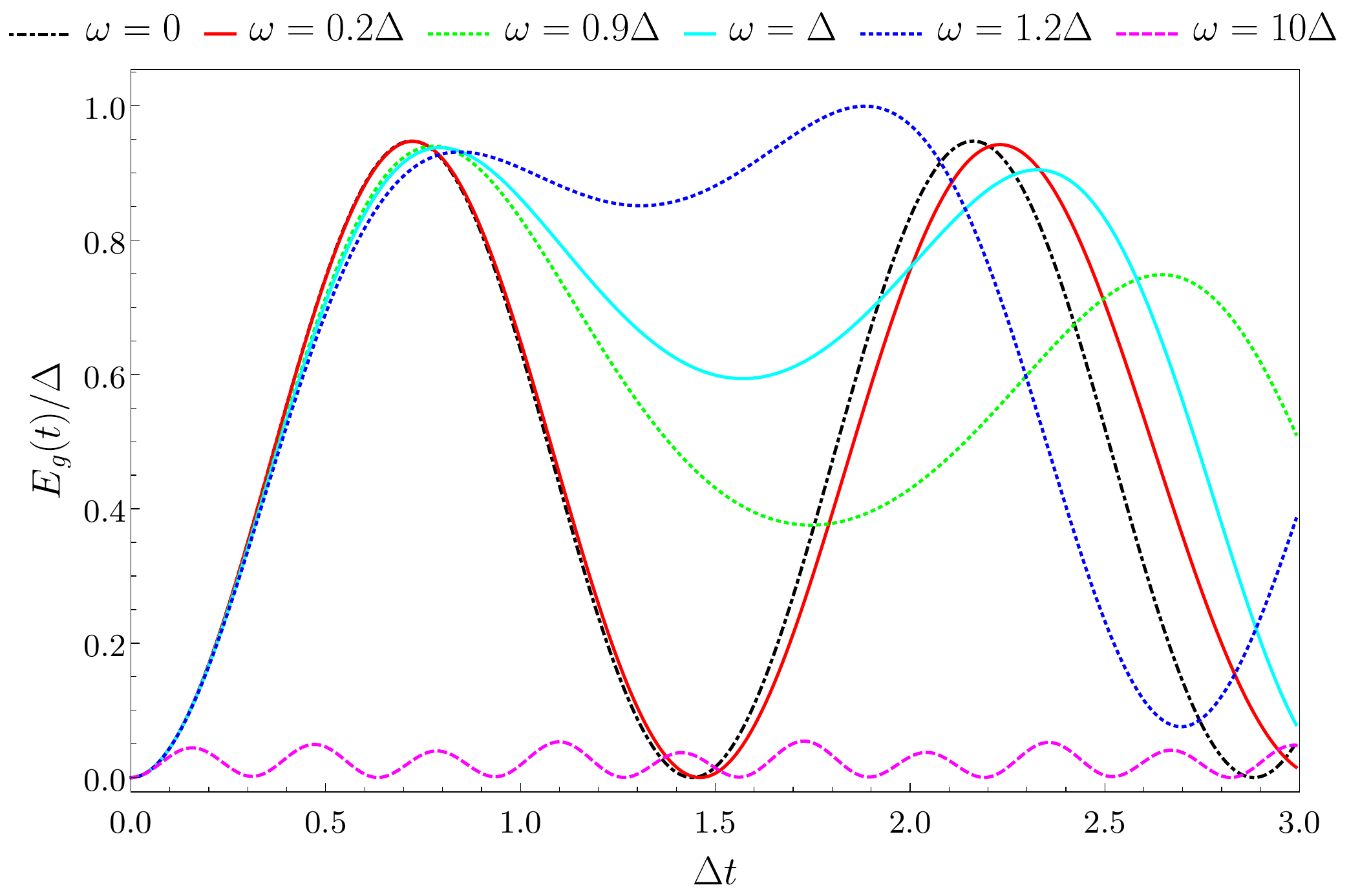}
\caption{Behaviour of $E_{g}(t)$ in units of $\Delta$ as a function of $\Delta t$ for: $\omega=0$ (dashed-dotted black curve), $\omega=0.2 \Delta$ (full red curve), $\omega=0.9\Delta$ (dotted green curve), $\omega=\Delta$ (full cyan curve), $\omega=1.2\Delta$ (dotted blue curve) and $\omega=10\Delta$ (dashed magenta curve). Curves are evaluated at fixed $A=3 \Delta$.}
\label{Figure5}
\end{figure}

In the previous discussion we have focused our attention only on a unique value of the frequency, namely $\omega=1.2 \Delta$.
We now analyze the behaviour of $E(t)$ for different external frequencies. To this end we discuss the representative case of the ground state for the cosine drive, but similar statements can be made for the other pure states and for the rectangular drive. In Figure \ref{Figure5} we report the dependence of $E_{g}(t)$ on the drive frequency $\omega$ (at fixed $A=3\Delta$).
We observe the evolution from a situation compatible with a static drive (dashed-dotted black and full red curves), to a situation close to the resonance (dotted green curve, full cyan curve, dotted blue curve), where the full charging is reached only for the latter case ($\omega=1.2\Delta$). By further increasing the frequency, we observe a strong suppression of the average energy stored (dashed green curve). This fact can be understood considering the asymptotic limit in Equation (\ref{Z_limit2}), (\ref{Z_limit21}), and (\ref{Z_limit22}) reported in \ref{app_c}, where we can notice that for all the initial conditions the energy for $\omega \gg A, \Delta$ decreases. From that expressions we can see that, at fixed drive amplitude, the ground state follows a $1/\omega^{2}$ power-law, while $|p_x\rangle$ and $|p_y\rangle$, drop as $1/\omega$.


\subsection{Energy quantum fluctuations}
\begin{figure}[htbp]
\centering
\includegraphics[width=1. \textwidth]{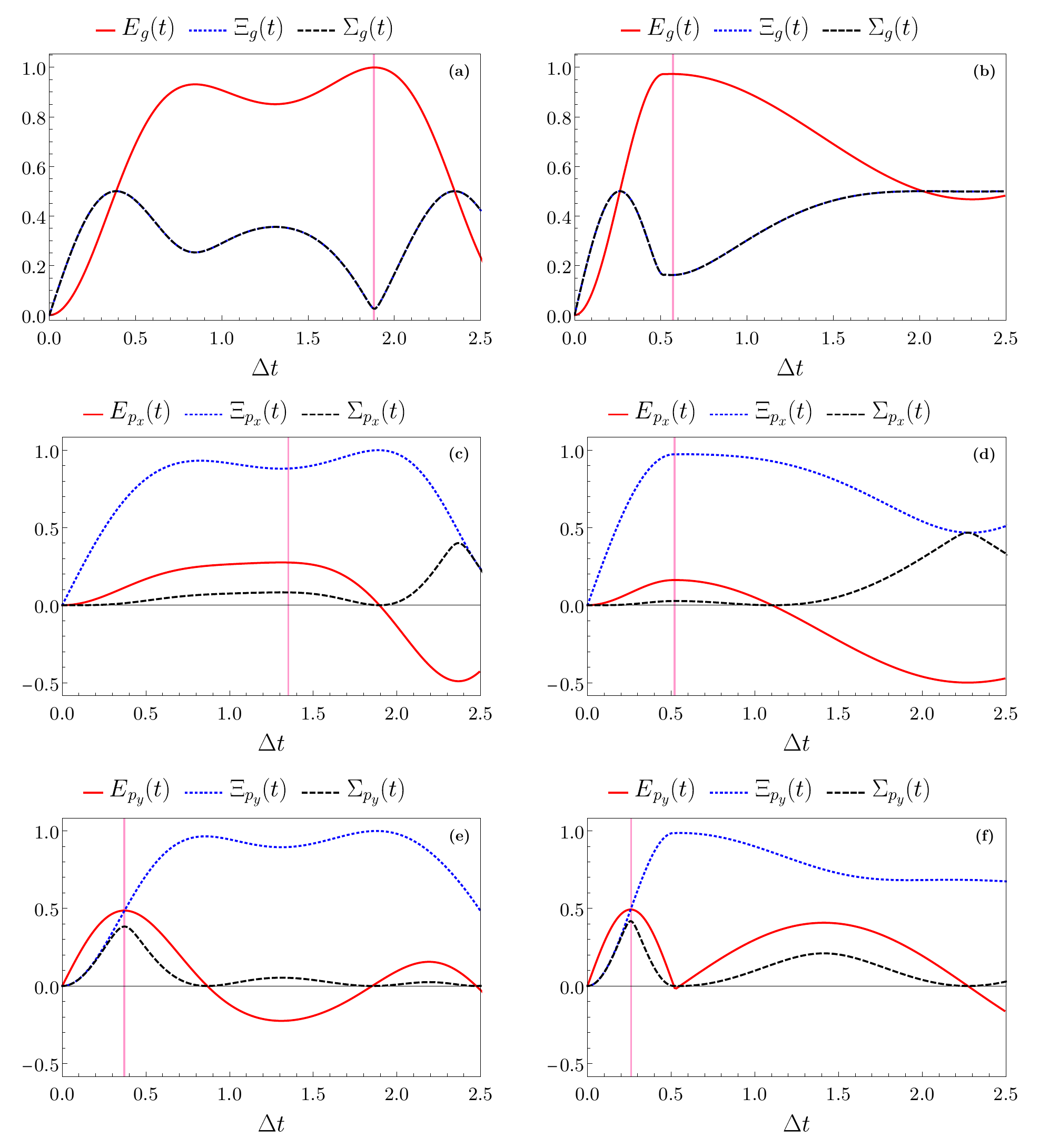}
\caption{Behaviour of $E(t)$ (full red curves), $\Xi(t)$ (dotted blue curves) and $\Sigma(t)$ (black dashed curves) in units of $\Delta$ as a function of $\Delta t$ at fixed $\omega=1.2\Delta$ and $A=3 \Delta$. Left panels represent the case of a cosine drive, while the right ones a train of rectangular pulses at $\eta=0.2$ choosing respectively as initial condition $|g\rangle$ a)-b), $|p_{x}\rangle$ c)-d) and $|p_{y}\rangle$ e)-f). Purple vertical lines determine the maxima of the stored energy.}
\label{Figure6}
\end{figure}

In Figure \ref{Figure6} we report the time behaviour of the average energy stored $E(t)$ together with the associated energy quantum fluctuations $\Xi(t)$ and $\Sigma(t)$, defined in Section \ref{Fluctuations}. 
 Here, we discuss both cosine and rectangular drives for initial conditions $|g\rangle$, $|p_{x}\rangle$ and $|p_{y}\rangle$. The parameters have been chosen as in the purple horizontal cuts of Figure \ref{Figure2} ($A=3\Delta$, $\omega=1.2\Delta$).
We now show that also energy quantum fluctuations strongly depend on the initial state preparation.  

For the initial condition in the ground state both fluctuations at equal times $\Xi$ and different times $\Sigma$ coincide for all possible drives (including of course the static case) due to the fact that $V_{g}(0)=0$ and $C_{g}(t)=0$ (see Equation (\ref{C}) evaluated at $P=1$), namely  
\begin{equation}
\Xi_{g}(t)=\Sigma_{g}(t)=\sqrt{V_{g}(t)}=\frac{\Delta}{2} \sqrt{1-\sigma^{2}_{z,g}(t)}.
\end{equation}
An interesting consequence of this relation is that, when the battery is completely charged ($E_{g}(t_{c})=\Delta$) at a given charging time  $t_{c}$, the two fluctuations are zero, leading to a ``noiseless'' charging process. Comparing Figure \ref{Figure6}a and Figure \ref{Figure6}b we can see that the parameters chosen for the comparison are ideal for the cosine drive, while they are not optimal for a rectangular drive. 

In particular the cosine drive approaches the full charging with negligible fluctuations. Instead, the chosen amplitude and the frequency of the drive are not ideal for the implementation of a QB in the case of a train of rectangular pulses. We will see shortly how the parameters can be improved in this case.

We now comment on the pure initial states $|p_x\rangle$ and $|p_y\rangle$. Here the two correlators have different behaviours.
In particular $\Xi(t)$ extends in the interval $0\leq \Xi\leq \Delta$, leading, in the time interval considered, to bigger fluctuations compared to $\Sigma(t)$ that varies in the interval $0\leq \Sigma\leq \Delta/2$. In addiction $\Sigma(t)$ is more related to $E(t)$, in particular their zeroes coincide. Moreover for the $|p_y\rangle$ state also the position of their first maxima coincide, meaning that the complete charging of the QB occurs at the time where the TLSs fluctuate the most, strongly affecting the potential use of the system. Instead, in the case of the $|p_x\rangle$ state the maximum of $\Sigma(t)$ coincides with the minimum of the energy and occurs at roughly $\Delta t\sim 2.3$ for both the drives.
Indeed, in correspondence of an ideal full charging we have\footnote{Notice that in Figure \ref{Figure6} e) and Figure \ref{Figure6} f) this value is not reached, since the condition $E_{p_y}(t_c)=\Delta/2$ is never fulfilled exactly neither for the cosine nor the rectangular drives, see Table \ref{tab1}.} 

\begin{equation}
\Sigma_{ p_x,  p_{y}}(t_{c,  p_x,  p_{y}})=\frac{\Delta}{2},
\end{equation}
resulting in strong fluctuation amplitude.
Considering the other correlator one has
\begin{equation}
\Xi_{ p_x, p_{y}}(t_{c,  p_x,  p_{y}})=\Delta \sqrt{\frac{1}{2}\bigg(\frac{1}{2}+\sigma_{z,g}(t_{c,  p_x,  {p_{y}}})\bigg)}.
\end{equation}

The last equation depends on the value of $\sigma_{z,g}$ at the full charging time calculated for the $|p_x\rangle$ and $|p_y\rangle$ state, leading to $\Xi_{p_x,p_{y}}(t_{c,p_x, p_{y}})=\Delta/2$ only if $\sigma_{z,g}(t_{c,{p_x,p_{y}}})=0$.
From Figure \ref{Figure6} c) and d) we can obviously see that this doesn't happen for the $|p_x\rangle$ state, where in the case of the cosine $\Xi(t_c,p_x)\sim0.88\Delta$ and for the rectangle $\Xi(t_c,p_x)\sim0.97\Delta$.

Instead we can observe that $\Xi_{p_{y}}(t_{c, p_{y}})\approx \Delta/2$ for both the cosine and the train of rectangular pulses, and the additional contribution present in the previous equation is thus small.
This is due to the fact that in the considered range of parameters $|\sigma^{(c)}_{z,g}(t_{c,{p_{y}}})|\ll 1/2$ ($\sigma^{(c)}_{z,g}(t_{c,{p_{y}}})=-0.02$ and $\sigma^{(r)}_{z,g}(t_{c,{p_{y}}})= 0.04$ respectively).From the above analysis we can argue that, as already observed discussing the averaged energy behaviour, the $|p_x\rangle$ initial state is not a good choice  for implementing a quantum battery. Moreover this state also shows great fluctuations in correspondence of the energy maximum. Unfortunately also the $|p_{y}\rangle$, although it shows a faster charging time $t_c$, compared to the ground state, it is subjected to unavoidable energy quantum fluctuations which would compromise its role has optimal initial state for a useful QB.
This picture holds true also for different values of the drive amplitude $A$. Moreover, by reducing the width of the peak of the rectangular pulse ($\eta=0.05$, $0.1$) the slower charging occurs together with greater quantum fluctuations. For the considered amplitude ($A=3 \Delta$) the static case, not shown, presents an incomplete charging which leads also to quantum fluctuations for all initial conditions, meaning that ac drives, in these conditions, are more suitable for the implementation of a QB.

Until now, we have examined parameters for which we could make comparisons between the two drives considered but that are not optimal for the train of rectangular pulses. Indeed, for the cosine the parameters $A=3\Delta$ and $\omega=1.2\Delta$ are close to the optimal choice because the relation $|\tilde{\Delta}|\gg \tilde{A}$ is well satisfied.

\begin{figure}[htbp]
\centering
\includegraphics[width=0.6 \textwidth]{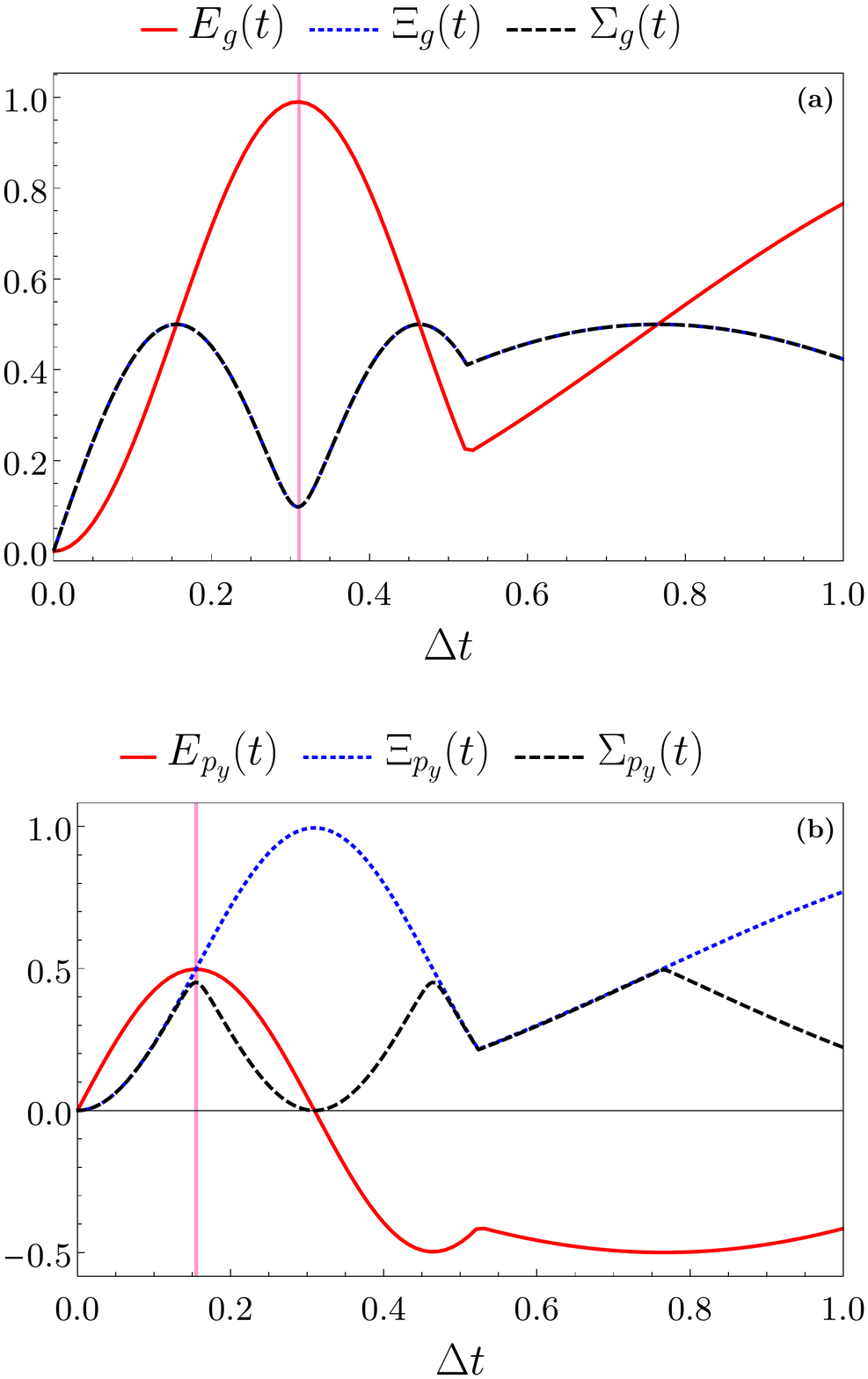}
\caption{Behaviour of $E(t)$ (full red curves), $\Xi(t)$ (dotted blue curves) and $\Sigma(t)$ black dashed curves) in units of $\Delta$ as a function of $\Delta t$ in the $|\tilde{\Delta}|\gg \tilde{A}$ regime. Here the parameters are fixed at $A=5 \Delta$ and $\omega=1.2\Delta$ for the rectangle drive at $\eta=0.2$ with initial condition: $|g\rangle$ a) and $|p_{y}\rangle$ b).}
\label{Figure7}
\end{figure}

In order to reach an optimal charging regime for the rectangular drive one possible choice of parameters is $A=5\Delta$ and $\omega=1.2\Delta$. This is reported in Figure \ref{Figure7}. Here we observe an almost complete charging of the battery for both the ground ($E_g(t) \approx 0.989\Delta$) and $|p_y\rangle$ ($E_{p_y}(t) \approx 0.497\Delta$) states.
These values are obtained also in faster times, i.e. $\Delta t^{(r)}_{c, g}\approx0.31$ and $\Delta t^{(r)}_{c, p_{y}}\approx0.16$. In this regime we also obtain less fluctuations in the ground state, as a consequence of the almost complete charging of the QB.

From this analysis we can conclude that the optimal drive amplitude for the rectangular pulse is higher with respect to the one for the cosine. Moreover, with a train of rectangular pulses it is possible to obtain a shorter charging time $t_c$. Ultimately it is a matter of the practical purpose one wants to pursue: using less power, related to the drive amplitude, and obtaining slightly longer charging process, or spending more power with a gain in the charging times.

\section{Conclusions}
\label{Conclusions}
In this paper, we have investigated a collection of $N$ independent two-level quantum systems coupled to a classical time dependent drive as an experimentally feasible example of quantum battery. We have investigated  performances of this system by means of exact numerical solution, showing comparison with analytical approximation within the so-called CHRW scheme which takes into account the effect of counter-rotating terms. Different preparation state (initial conditions) and shapes of the external driving have been analyzed and discussed.
As useful figures of merit for the QB, we have studied the average energy stored and also its quantum fluctuations. The latter has been considered by inspecting the behaviour of two correlators, at equal and different times, during the charging protocol.
The main finding of our analysis is the fact that a charging protocol starting from a completely empty battery (ground state of the two-level system) leads to an optimal charging in absence of energy fluctuations. Other possible initial states are either more affected by fluctuations ($|p_{y}\rangle$) or characterized by a longer charging time ($|p_{x}\rangle$), as a consequence of the path followed by the state evolution on the surface of the Bloch sphere. Moreover, we identified a range of parameters where a train of peaked rectangular pulses leads to a faster charging with respect to both the usually investigated harmonic and the static case.

\section*{Acknowledgments}
M.C. acknowledges support from the Quant-EraNet project ``Supertop". Authors would like to thanks M. Acciai for useful discussions.


\appendix

\section{Analysis of mixed states as initial state}
\label{app_a}

In the main text, we have considered only pure initial states of the TLS because they allow to reach the best performances of the QB. 
Here, to further strengthen this statement, we want to analyze other intermediate states of the form

\begin{equation} |\psi(0)\rangle= \gamma|g\rangle+\delta|e\rangle, \end{equation}

where $\gamma, \delta \in \mathbb{C}$ and $|\gamma|^2+|\delta|^2=1$.

In particular  we consider the representative examples

\begin{equation} |p_\gamma\rangle= \frac{1}{2}(|g\rangle+\sqrt{3}| e\rangle) \end{equation}

where $\gamma=1/2$ and $\delta=\sqrt{3}/2$, and

\begin{equation} |p_{\gamma'}\rangle= \frac{1}{2}(\sqrt{3}|g\rangle+| e\rangle) \end{equation}

where $\gamma=\sqrt{3}/2$ and $\delta=1/2$. 
We recall that, being the initial density matrix in the form of Equation (\ref{rho_zero}), here we have $P_\gamma=1/4$ and $P_{\gamma'}=3/4$, $\alpha=\sqrt{3}/4$ and $\beta=0$.

\begin{figure}[htbp]
\centering
\includegraphics[width=0.95 \textwidth]{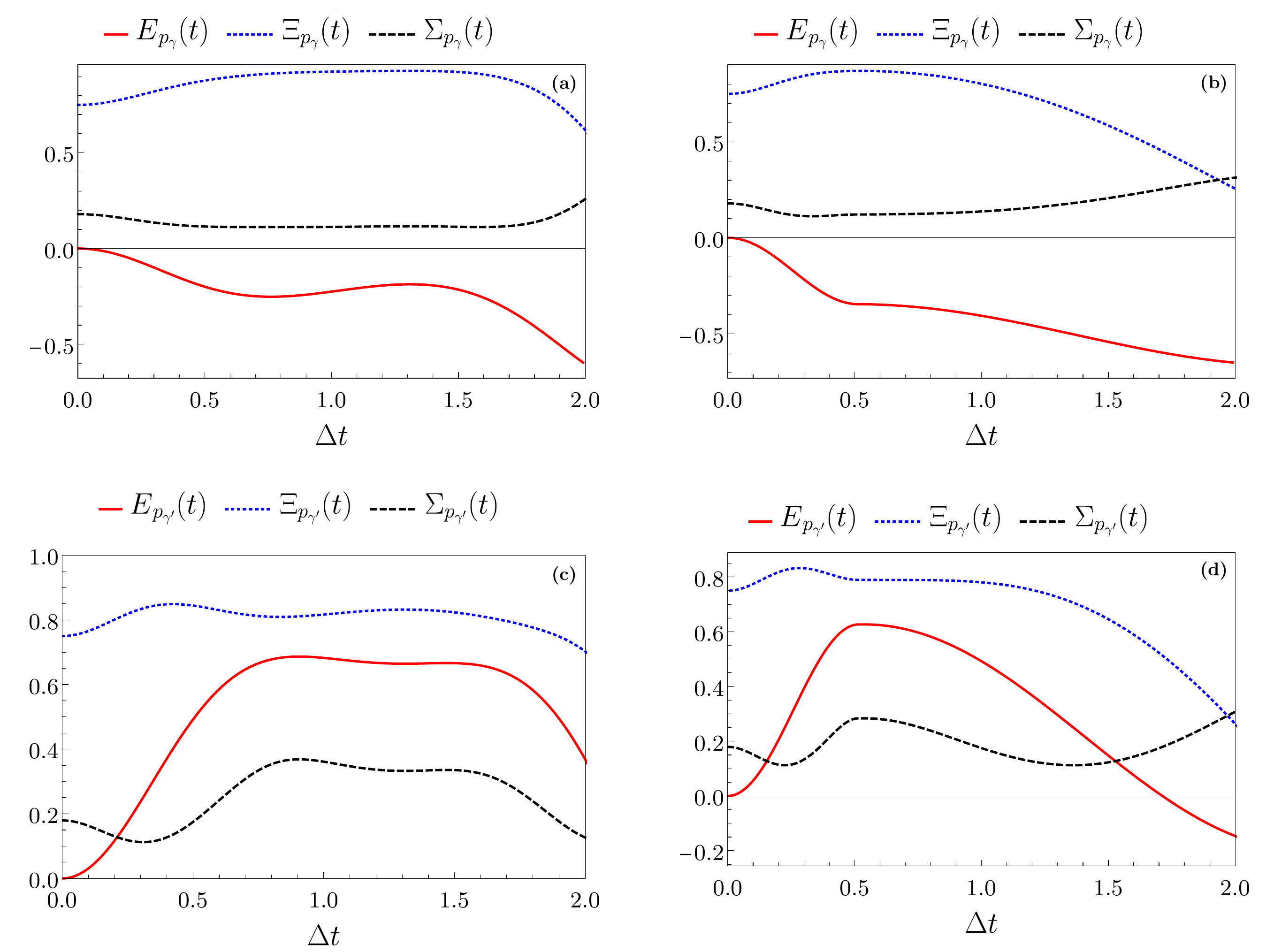}
\caption{Behaviour of $E(t)$ (red curves), $\Xi(t)$ (blue dotted curves) and $\Sigma(t)$ (black dash-dotted curves) in unit of $\Delta$ for the cosine drive a) and c) and for the rectangular pulse ($\eta=0.2$) b) and d). Panel a) and b) represent the state $|p_{\gamma}\rangle$ while panel c) and d) represent $|p_{\gamma'}\rangle$. Other parameters are $A=3\Delta$ and $\omega=1.2\Delta$.}
\label{Figure8}
\end{figure}

In Figure \ref{Figure8} we report the behaviour of $E(t)$ (red curves), $\Xi(t)$ (blue dotted curves) and $\Sigma(t)$ (black dash-dotted curves) in unit of $\Delta$ for the cosine drive in panel a) and c) and for the rectangular pulses ($\eta=0.2$) in panel b) and d), for $A=3\Delta$ and $\omega=1.2\Delta$. 

Here we note that both the chosen initial states $ |p_\gamma\rangle$ and $|p_{\gamma'}\rangle$ never reach the full charging of the battery. In particular, with the state $|p_\gamma\rangle$, it is only possible to discharge the battery, while for $|p_{\gamma'}\rangle$ in the case of the cosine we have a maximum of the energy $E \approx 0.686\Delta$ at the time $\Delta t_c^{(c)} \approx0.91$ and in the case of the rectangular pulse we obtain $E \approx 0.626\Delta$ at the time $\Delta t_c ^{(r)}\approx0.56$, meaning that we never reach the full charge of the QB.

From the curves in Figure \ref{Figure8} we can also see that unavoidable fluctuations are present in the system.
This means that even though $|p_{\gamma'}\rangle$ can achieve a good charging of the battery, large fluctuations are present, compared to the ground state. 

We now consider mixed states, where $\alpha=\beta=0$ and $P\neq 0,1$. Notice that, for $P>1/2$ these can be considered as a prototype for thermal states according to the relation

\begin{equation} P=e^{\Delta/2k_BT}/Z \end{equation}

with $Z=e^{\Delta/2k_BT}+e^{-\Delta/2k_BT}$, $k_B$ the Boltzmann constant, $T$ an effective temperature and $\Delta$ the level spacing between the ground state and the excited state of the TLS.

To use the equations in Section \ref{Fluctuations}, we need to write the energy of the QB in these states, using Equation (\ref{Energy}), as 

\begin{equation}\label{Emix} E_m(t)= \frac{\Delta}{2}(2P-1)E_g(t), \end{equation}

where the index $m$ indicates the mixed states. We can now proceed analyzing two different initial states, namely the one with $P_m=1/4$ and $P_{m'}=3/4$ that herafter we refer to as $P_m$ and $P_{m'}$.

\begin{figure}[htbp]
\centering
\includegraphics[width=0.95 \textwidth]{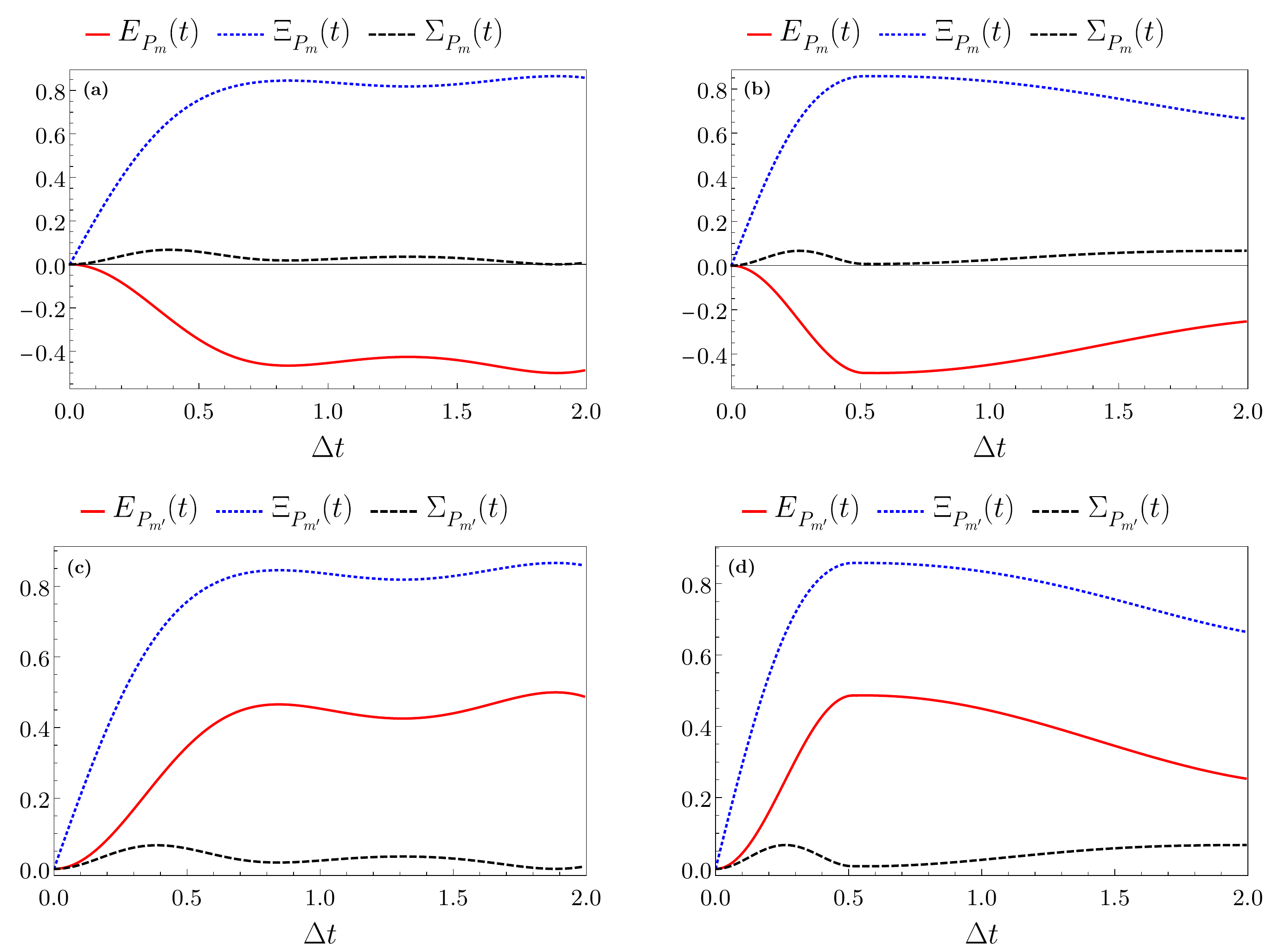}
\caption{Behaviour of $E(t)$ (red curves), $\Xi(t)$ (blue dotted curves) and $\Sigma(t)$ (black dash-dotted curves) in unit of $\Delta$ for the cosine drive a) and c) and for the rectangular pulse ($\eta=0.2$) b) and d). Panel a) and b) represent the state $P_m$ while panel c) and d) represent $P_{m'}$. Other parameters are $A=3\Delta$ and $\omega=1.2\Delta$.}
\label{Figure9}
\end{figure}

In Figure \ref{Figure9} we report the behaviour of $E(t)$ (red curves), $\Xi(t)$ (blue dotted curves) and $\Sigma(t)$ (black dash-dotted curves) in unit of $\Delta$ for the cosine drive in panel a) and c) and for the rectangular pulses ($\eta=0.2$) in panel b) and d), for $A=3\Delta$ and $\omega=1.2\Delta$.
From the curves of the average energy we can see that these initial mixed states behave differently. In particular $P_m$ (panels a) and b)) is a passive state from which we can only subtract energy while $P_{m'}$ (panels c) and d)) is an active state to which we can only provide energy (see description at the end of Section \ref{Fluctuations}). 
In general we can see that none of these cases are optimal.
  
This can be demonstrated looking at Equation (\ref{Emix}), where choosing the maximum ideal value of $E_g(t)=\Delta$ implies that

\begin{equation} E_m(t)=(2P-1)\Delta. \end{equation}

Since mixed states have $0< P < 1$, the battery will never reach $E_m(t)=\pm\Delta$ as can be seen in Figure \ref{Figure9}, where 
panels a) and b) show an incomplete discharge, and panels c) and d) an incomplete charge of the battery.

Moreover, these states also lead to big fluctuations, concerning mostly the correlator $\Xi(t)$, which make mixed states not optimal to build a QB. We also note that the fluctuations for the two different states are identical, this is because in Equations (\ref{W_compact}) and (\ref{Sigma_compact}) the contributions are of the form

\begin{eqnarray}
V_m(t)&=& \frac{\Delta^2}{4}\bigg[1-(2P-1)^2\sigma_{z,g}^2\bigg]\\
V_m(0)&=& \frac{\Delta^2}{4}\bigg[1-(2P-1)^2\bigg]\\
C_m(t)&=&\frac{\Delta^2}{2}\bigg[(2P-1)^2-1\bigg]\sigma_{z,g}.
\end{eqnarray}

Here for the chosen values of $P$, we have that $(2P-1)^2=1/4$.

The above analysis confirms the statement in the main text (Section \ref{pis}), in which we explain our decisions to use three different pure states as initial sates for the TLS.


\section{Time evolution of the considered initial states in the CHRW approximation}
\label{app_b}
To derive Equations (\ref{Z_g}), (\ref{Z_px}) and (\ref{Z_py}) in the main text it is useful to write explicitly the time evolved state $\psi(t)$ in the CHRW approximation for each initial state considered in Section \ref{Initial}. To do so we evaluate the time evolution of the initial state according to 
\begin{equation}
\label{timeev} 
|\psi(t)\rangle=\mathcal{U}^\dagger(t)\mathcal{S}^\dagger(t)e^{-i\tilde{\mathcal{H}}t}|\psi(0)\rangle 
\end{equation}
where $\tilde{\mathcal{H}}$ is the Hamiltonian in Equation (\ref{H_final}), and we applied the unitary transformations in Equations (\ref{U(t)}) and (\ref{S(t)}).

The time evolution of the ground state $|g\rangle$ according to the CHRW approximation is given by
\begin{eqnarray}
\label{phi_g} 
|\psi_{g}(t) \rangle&=& \left\{\cos[ \pi \bar{z} \varphi(t)]e^{i\frac{\omega t}{2}}\left[\cos\left( \frac{\Omega_R}{2} t\right)  + i\left(\frac{ \tilde{\Delta}}{\Omega_R}\right)\sin\left( \frac{\Omega_R}{2}  t\right) \right]\right.\nonumber\\
&-&\left. \left(\frac{2\tilde{A}}{\Omega_R}\right)\sin \left(\frac{\Omega_R}{2}t\right)  \sin  [\pi \bar{z} \varphi(t)]e^{-i\frac{\omega t}{2}} \right\}|g\rangle \nonumber \\
&+&  \left\{\sin [ \pi \bar{z} \varphi(t)] e^{i\frac{\omega t}{2}}\left[-i \cos \left(\frac{\Omega_R}{2} t\right) + \left(\frac{ \tilde{\Delta}}{\Omega_R}\right)\sin\left(\frac{\Omega_R}{2} t\right) \right]\right.\nonumber\\ 
&-&\left. 2i\left(\frac{\tilde{A}}{\Omega_R}\right) \sin\left(\frac{\Omega_R}{2}t \right) \cos  [ \pi \bar{z} \varphi(t)]e^{-i\frac{\omega t}{2}} \right\}  |e\rangle. \nonumber \\
\end{eqnarray}

Analogously the dynamics of $|p_x\rangle$ in the CHRW approximation reads 
\begin{eqnarray}
\label{phi_px}
|\psi_{p_x}(t)\rangle&=&  \left\{\cos [ \pi \bar{z} \varphi(t)] e^{i\frac{\omega t}{2}}\left[\cos\left(\frac{\Omega_R}{2}t\right)+i \left(\frac{\tilde{\Delta}-2\tilde{A}}{\Omega_R}\right)\sin\left(\frac{\Omega_R}{2}t\right)\right]\right.\nonumber\\
&-&\left.\sin[ \pi \bar{z} \varphi(t)] e^{-i\frac{\omega t}{2}}\left[\left(\frac{\tilde{\Delta}+2\tilde{A}}{\Omega_R}\right)\sin\left(\frac{\Omega_R}{2}t\right)-i\cos\left(\frac{\Omega_R}{2}t\right) \right] \right\}|g\rangle  \nonumber\\
&+&\left\{\cos[ \pi \bar{z} \varphi(t)]  e^{-i\frac{\omega t}{2}}\left[\cos\left(\frac{\Omega_R}{2}t\right) -i\left(\frac{\tilde{\Delta}+2\tilde{A}}{\Omega_R}\right)\sin\left(\frac{\Omega_R}{2}t\right)\right] \right.\nonumber\\
&+& \left.\sin[ \pi \bar{z} \varphi(t)]  e^{i\frac{\omega t}{2}}\left[\left(\frac{\tilde{\Delta}-2\tilde{A}}{\Omega_R}\right)\sin\left(\frac{\Omega_R}{2}t\right)-i\cos\left(\frac{\Omega_R}{2}t\right)\right] \right\} |e\rangle.\nonumber \\ 
\end{eqnarray}

Finally, the state $|p_y\rangle $ evolves as
\begin{eqnarray}
\label{phi_py}
|\psi_{p_y}(t)\rangle&=& \left\{\cos[\pi \bar{z} \varphi(t)]  e^{i\frac{\omega t}{2}}\left[\cos\left(\frac{\Omega_R}{2}t\right)+\left(\frac{i\tilde{\Delta}-2\tilde{A}}{\Omega_R}\right)\sin\left(\frac{\Omega_R}{2}t\right)\right]\right.\nonumber\\
&+&\left.\sin[\pi \bar{z} \varphi(t)]  e^{-i\frac{\omega t}{2}}\left[\left(\frac{i\tilde{\Delta}-2\tilde{A}}{\Omega_R}\right)\sin\left(\frac{\Omega_R}{2}t\right)-\cos\left(\frac{\Omega_R}{2}t\right) \right] \right\}  |g\rangle\nonumber\\
&+& \left\{\cos[\pi \bar{z} \varphi(t)]  e^{-i\frac{\omega t}{2}}\left[-\left(\frac{\tilde{\Delta}+2i\tilde{A}}{\Omega_R}\right)\sin\left(\frac{\Omega_R}{2}t\right)-i\cos\left(\frac{\Omega_R}{2}t\right)\right]\right.\nonumber\\
&+&\left.\sin[\pi \bar{z} \varphi(t)]  e^{i\frac{\omega t}{2}}\left[-i\cos\left(\frac{\Omega_R}{2}t\right)+\left(\frac{\tilde{\Delta}+2i\tilde{A}}{\Omega_R}\right)\sin\left(\frac{\Omega_R}{2}t\right)\right] \right\} |e\rangle. \nonumber \\
\end{eqnarray}

In the above Equations we have considered the definition for $\varphi(t)$, $\tilde{A}$, $\tilde{\Delta}$ and $\Omega_R$ in Equations (\ref{varphi}), (\ref{atilde}), (\ref{dtilde}) and (\ref{Rabi}) respectively. In addiction $\bar{z}$ is the parameter that solve Equation (\ref{bar_chi}).

\section{Asymptotic regimes for the stored energy in the CHRW approximation}
\label{app_c}

\begin{figure}[htbp]
\centering
\includegraphics[width=1 \textwidth]{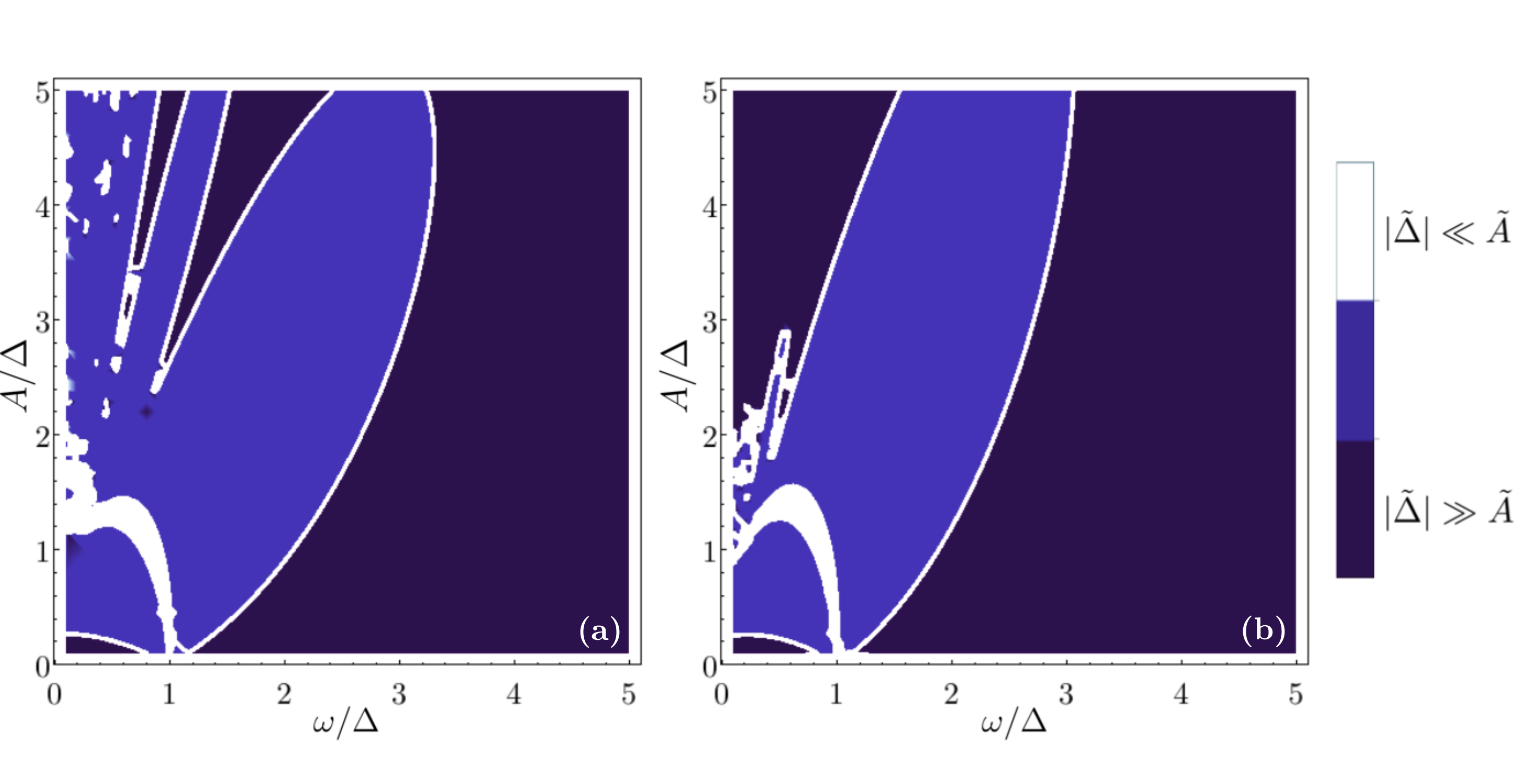}
\caption{Density plot of $\tilde{A} / |\tilde{\Delta}| $ as a function of $A$ and $\omega$ for the cosine drive a) and the rectangular drive ($\eta=0.2$) b). Dark blue regions represent $|\tilde{\Delta}| \gg \tilde{A}$, white regions represent $|\tilde{\Delta}| \ll \tilde{A}$, while the violet regions indicate all the other cases.}
\label{Figure10}
\end{figure}

Some useful limits are worth to be discussed for the average energy stored, derived in Section \ref{Relevant}. 
First we consider the regime $|\tilde{\Delta}| \gg \tilde{A}$ (see the dark blue regions in Figure \ref{Figure10}), in this case the renormalized frequency in Equation (\ref{Rabi}) reduces to $\Omega_{R} \approx |\tilde{\Delta}|$ and 
\begin{eqnarray}
\label{Z_limit1_g}
\frac{E_{g}(t)}{\Delta}&\approx&\frac{1}{2}\left\{1 -\cos\left[2 \pi \bar{z} \varphi(t) \right]\right\}\\
\label{Z_limit1_px}
\frac{E_{p_{x}}(t)}{\Delta}&\approx& \frac{1}{2}\sin\left[2 \pi \bar{z} \varphi(t)\right] \sin \left[\Delta p_0(\bar{z})t\right] \\
\label{Z_limit1_py}
\frac{E_{p_{y}}(t)}{\Delta}&\approx& \frac{1}{2}\sin\left[2\pi \bar{z} \varphi(t)\right] \cos\left[ \Delta p_0(\bar{z})t \right],
\end{eqnarray}
where $p_0$ is the zeroth order photoassisted coefficient obtained from Equation (\ref{p_l}) at $l=0$.
Notice that in this limit the dynamics induced by the Hamiltonian in (\ref{H_final}) becomes very simple due to the fact that only the diagonal terms survive. Given the above expressions we can evaluate the time $t_c$ at which the QB reach the full charging for different initial states and drives. 
For the ground state we obtain the following equation
\begin{equation}\label{tcg} 
2\pi \bar{z} \varphi(t_{c})= (2n+1)\pi, 
\end{equation}
where $n \in \mathbb{Z}$ and $\bar{z}$ solve Equation (\ref{bar_chi}) univocally for a fixed couple of parameters ($A/\Delta$, $\omega/\Delta$), provided to chose them in the proper regime of validity of the CHRW approximation (see Section \ref{Results}). Notice that, because the action of the external drive starts at $t=0$ (see Equation (\ref{H})), we only consider as meaningful charging times such as $t_{c}>0$ and for actual practical purposes we are interested in selecting the shorter among them. 

Starting from the $|p_x\rangle$ state, since the product of the two terms in Equation (\ref{Z_limit1_px}) need to be equal to $1$ to have a maximum of the energy, we need to fulfill

\begin{eqnarray}\label{tcpx}  \sin (\Delta p_0(\bar{z})t_{c})=\pm 1 \Rightarrow \Delta t_{c}= \frac{(2n+1)\pi}{2p_0(\bar{z})},
\end{eqnarray}
with $n \in \mathbb{Z} $. Consequently one needs
\begin{eqnarray} 
\sin [2\pi\bar{z}\varphi(t_{c})]=\pm 1. 
\end{eqnarray}

The same consideration can be done for the $|p_y\rangle$ state

\begin{eqnarray}\label{tcpy} 
\cos \Delta p_0(\bar{z})t_{c}=\pm1 \Rightarrow \Delta t_{c}= \frac{n\pi}{p_0(\bar{z})},
\end{eqnarray}
where $n \in \mathbb{Z}$, and consequently one needs
\begin{equation} 
\sin \left[2\pi\bar{z}\varphi(t_{c})\right]=\pm1. 
\end{equation}
In both cases it is not always possible to find solutions for the system of equations. Indeed, we need to chose a pair ($A/\Delta$, $\omega/\Delta$) leading to a fixed $\bar{z}$ and able to fulfill the first and the second equation. 

The opposite case  $|\tilde{\Delta}| \ll \tilde{A}$ is represented by the white regions in Figure \ref{Figure10} (see \cite{Zhang19, Chen19} for the $\tilde{\Delta}=0$ case with ground state initial condition and in presence of a harmonic drive). In this limit the Hamiltonian in (\ref{H_final}) is off-diagonal and the expressions for the stored energies read
\begin{eqnarray}
\label{E_gApp}
\frac{E_{g}(t)}{\Delta} &=& \frac{1}{2}\left[-\cos(2\tilde{A} t)\cos\left[2\pi \bar{z} \varphi(t)\right]+\sin(2\tilde{A} t)\sin\left[2\pi \bar{z} \varphi(t)\right]\cos(\omega t)+ 1\right]\nonumber\\
\\
\label{E_pxApp}
\frac{E_{p_{x}}(t)}{\Delta} &=& \frac{1}{2}\sin\left[2\pi \bar{z} \varphi(t)\right]\sin\left(\omega t\right)\\
\label{E_pyApp}
\frac{E_{p_{y}}(t)}{\Delta} &=& \frac{1}{2}\left[\sin(2\tilde{A} t)\cos\left[2\pi \bar{z} \varphi(t)\right] +\cos(2\tilde{A} t)\sin\left[2\pi \bar{z} \varphi(t)\right]\cos\left(\omega t\right)\right].\nonumber \\
\end{eqnarray}
Here, simple analytic forms for $t_c$ can be derived in the case in which we turn on the external drive for exactly one period $t_{c}=T=2\pi/\omega$ in the case of the ground state \cite{Zhang19} and $|p_y\rangle$ state. In the first case we have

\begin{eqnarray} \sin^2 (\tilde{A}T)=1  \Rightarrow T=\frac{(2n+1)\pi}{2\tilde{A}}, \end{eqnarray}
where $n \in \mathbb{Z}$. Similarly for $|p_y\rangle$

\begin{eqnarray} \frac{1}{2}\sin (2\tilde{A}T)=\frac{1}{2}  \Rightarrow T=\frac{(4n+1)\pi}{4\tilde{A}}, \end{eqnarray}

where $n \in \mathbb{N}$. Finally for the $|p_x\rangle$ state we obtain a simple form for $t_c$ when we consider half a period of the external drive, namely

\begin{eqnarray}  \frac{1}{2}\sin \bigg[2\pi\bar{z}\varphi\bigg(\frac{T}{2}\bigg)\bigg]  \Rightarrow \varphi\bigg(\frac{T}{2}\bigg)=\frac{(4n+1)\pi}{4\bar{z}}, \end{eqnarray}
where $n \in \mathbb{N}$. Here we can see that the value of the time of charging can be written explicitly only specifying the drive. 

In the $\omega \ll \Delta$ our expressions map into the static case (denoted with the index $s$) in presence of a constant external bias of amplitude $A$. It can be recovered from the previous results setting $\varphi(t)=0$ (in this case the drive is purely DC) and replacing $\tilde{\Delta}\rightarrow \Delta$, $\tilde{A}\rightarrow A/2$. In this case we clearly observe a unique characteristic frequency for the system, namely the bare Rabi frequency $\Omega^{(s)}_{R}=\sqrt{\Delta^{2}+A^{2}}$ with stored energy evolving in time as 
\begin{eqnarray}
\label{E_g_static}
\frac{E^{(s)}_{g}(t)}{\Delta} &=&\frac{1}{2}\frac{A^{2}}{A^{2}+\Delta^{2}} \left[1-\cos(\Omega_{R}^{(s)} t)\right]\\
\label{E_px_static}
\frac{E^{(s)}_{p_{x}}(t)}{\Delta} &=& \frac{1}{2}\frac{A \Delta}{A^{2}+\Delta^{2}} \left[1-\cos(\Omega_{R}^{(s)} t)\right]\\
\label{E_py_static}
\frac{E^{(s)}_{p_{y}}(t)}{\Delta} &=&\frac{1}{2}\frac{A}{\sqrt{A^{2}+\Delta^{2}}} \sin(\Omega_{R}^{(s)} t).
\end{eqnarray}

Conversely in the high frequency limit $\omega\gg \Delta, A$ one has
\begin{eqnarray}
\label{Z_limit2}
\frac{E_{g}(t)}{\Delta}&\approx& \frac{1}{2}\left\{1-\cos\left[2 \pi \bar{z} \varphi(t)\right]\right\}\propto \frac{1}{\omega^{2}}\\
\label{Z_limit21}
\frac{E_{p_{x}}(t)}{\Delta}&\approx&\frac{1}{2}\sin\left[2\pi \bar{z} \varphi(t)\right]\propto\frac{1}{\omega}  \\
\label{Z_limit22}
\frac{E_{p_{y}}(t)}{\Delta}&\approx& \frac{1}{2}\sin\left[2\pi \bar{z} \varphi(t)\right]\propto\frac{1}{\omega}. 
\end{eqnarray}
This power-law decay is a consequence of the fact that in this limit $p_1(\bar{z})-p_{-1}(\bar{z}) \rightarrow 0$ and $\bar{z} \rightarrow A/\omega$. This means that all the $\sigma_{z}(t)$ approaches $0$ by increasing $\omega$ (at fixed value of $A$) making impossible to fulfill the condition $\bar{z} \varphi(t)=1$.


\end{document}